\newif{\ifchangetext}
\changetexttrue

\documentclass[iop]{emulateapj}
\newcommand{\cmt}[1]{}
\ifchangetext

\else

\fi

\usepackage{natbib}   
\usepackage{amsmath}  
\usepackage{verbatim} 
\usepackage{enumerate}
\usepackage{amssymb}  
\usepackage[colorlinks=true,citecolor=blue,linkcolor=blue,urlcolor=blue]{hyperref}      
\usepackage[usenames,dvipsnames]{xcolor}  
\usepackage{appendix}

\DeclareGraphicsExtensions{.pdf,.png,.jpg}

\defcitealias{Rodney:2014}{R14}
\def\R14{\citetalias{Rodney:2014}}

\defcitealias{Graur:2014}{G14}
\def\G14{\citetalias{Graur:2014}}

\shorttitle{CANDELS + CLASH SN Cosmology}
\shortauthors{Riess et al.}

\begin{document}

\title{Type Ia Supernova Distances at Redshift $> 1.5$ from the Hubble Space Telescope Multi-Cycle Treasury Programs: The Early Expansion Rate}


\newcommand{\JHU}{Department of Physics and Astronomy, The Johns Hopkins University, 3400 North Charles Street, Baltimore, MD 21218, USA}
\newcommand{\Pitt}{Department of Physics and Astronomy, University of Pittsburgh, Pittsburgh, PA 15260, USA}
\newcommand{\STScI}{Space Telescope Science Institute, 3700 San Martin Drive, Baltimore, MD 21218, USA}
\newcommand{\Chicago}{Department of Physics, University of Chicago, Chicago, IL 60637, USA}
\newcommand{\KICP}{Kavli Institute for Cosmological Physics, University of Chicago, 5640 South Ellis Avenue, Chicago, IL 60637}
\newcommand{\Berkeley}{Department of Astronomy, University of California, Berkeley, CA 94720-3411, USA}
\newcommand{\BerkeleyPhys}{Department of Physics, University of California, Berkeley, CA 94720, USA}
\newcommand{\LBNL}{Lawrence Berkeley National Laboratory, Berkeley, CA 94720, USA}
\newcommand{\Riverside}{Department of Physics and Astronomy, University of California, 900 University Avenue, Riverside, CA 92521, USA}
\newcommand{\SaoPaulo}{Instituto de Astronomia, Geof\'isica e Ci\^encias Atmosf\'ericas, Universidade de S\~ao Paulo, Cidade Universit\'aria, 05508-090, S\~ao Paulo, Brazil}
\newcommand{\Andalucia}{Instituto de Astrof\'isica de Andaluc\'ia (CSIC), E-18080 Granada, Spain}
\newcommand{\Cantabria}{IFCA, Instituto de F\'isica de Cantabria (UC-CSIC), Av. de Los Castros s/n, 39005 Santander, Spain}
\newcommand{\WKU}{Department of Physics, Western Kentucky University, Bowling Green, KY 42101, USA}
\newcommand{\AMNH}{Department of Astrophysics, American Museum of Natural History, Central Park West and 79th Street, New York, NY 10024, USA}
\newcommand{\NYU}{Center for Cosmology and Particle Physics, New York University, New York, NY 10003, USA}
\newcommand{\Copenhagen}{Dark Cosmology Centre, Niels Bohr Institute, University of Copenhagen, Juliane Maries Vej 30, DK-2100 Copenhagen, Denmark}
\newcommand{\DARK}{Dark Cosmology Centre, Niels Bohr Institute, University of Copenhagen, Juliane Maries Vej 30, DK-2100 Copenhagen, Denmark}
\newcommand{\Arizona}{Department of Astronomy, University of Arizona, Tucson, AZ 85721, USA}
\newcommand{\UCSC}{Department of Astronomy and Astrophysics, University of California, Santa Cruz, CA 95064, USA}
\newcommand{\NotreDame}{Department of Physics, University of Notre Dame, Notre Dame, IN 46556, USA}
\newcommand{\TelAviv}{School of Physics and Astronomy, Tel Aviv University, Tel Aviv 69978, Israel}
\newcommand{\Rutgers}{Department of Physics and Astronomy, Rutgers, The State University of New Jersey, Piscataway, NJ 08854, USA}
\newcommand{\CfA}{Harvard-Smithsonian Center for Astrophysics, 60 Garden Street, Cambridge, MA 02138, USA}
\newcommand{\Minnesota}{Department of Astronomy, University of Minnesota, 116 Church Street SE, Minneapolis, MN 55455, USA}
\newcommand{\NOAO}{National Optical Astronomy Observatory, 950 North Cherry Avenue, Tucson, AZ 85719, USA}
\newcommand{\UCSB}{Department of Physics, University of California, Santa Barbara, CA 93106, USA}
\newcommand{\LCOGT}{Las Cumbres Observatory Global Telescope Network, 6740 Cortona Drive, Goleta, CA 93117, USA}
\newcommand{\Colby}{Colby College, 4000 Mayflower Hill Drive, Waterville, ME 04901, USA}
\newcommand{\Kentucky}{University of Kentucky, Lexington, KY 40506}
\newcommand{\UCDavis}{University of California, Davis, 1 Shields Avenue, Davis, CA 95616}
\newcommand{\Irvine}{Department of Physics and Astronomy, University of California, Irvine, CA 92697, USA}
\newcommand{\Steward}{Steward Observatory, University of Arizona, 933 North Cherry Avenue Tucson, AZ 85721, USA}
\newcommand{\GSFC}{Astrophysics Science Division, NASA Goddard Space Flight Center, Mail Code 661, Greenbelt, MD 20771, USA}
\newcommand{\JSSI}{Joint Space Science Institute, University of Maryland, College Park, Maryland 20742, USA}
\newcommand{\ESO}{European Southern Observatory, Garching bei M\"unchen, Germany}
\newcommand{\ECUmunich}{Excellence Cluster Universe, Technische Universit\"at M\"unchen, Germany}
\newcommand{\PSU}{Department of Astronomy and Astrophysics, Pennsylvania State University, University Park, PA 16802, USA}
\newcommand{\UT}{Department of Astronomy, University of Texas, Austin, TX 78712, USA}
\newcommand{\ASU}{School of Earth and Space Exploration, Arizona State University, Tempe, AZ 85287, USA}
\newcommand{\UK}{Department of Physics and Astronomy, University of Kentucky, Lexington, KY 40506, USA}
\newcommand{\UConn}{Department of Physics, University of Connecticut, 2152 Hillside Road, Storrs, CT 06269}
\newcommand{\USC}{Department of Physics and Astronomy, University of South Carolina, 712 Main Street, Columbia, SC 29208, USA}
\newcommand{\UPenn}{Department of Physics and Astronomy, University of Pennsylvania, Philadelphia, PA 19104, USA}
\newcommand{\IPAC}{Infrared Processing and Analysis Center, California Institute of Technology, Pasadena, CA 91125, USA}

\newcommand{\HubbleFellow}{Hubble Fellow}
\newcommand{\NSF}{NSF Astronomy and Astrophysics Postdoctoral Fellow}
\newcommand{\ChancellorFellow}{University of California Chancellor's Postdoctoral Fellow}
\newcommand{\MillerFellow}{Miller Senior Fellow, Miller Institute for Basic Research in Science, University of California, Berkeley, CA 94720, USA}

\newcounter{affilct}
\setcounter{affilct}{0}

\makeatletter
\newcommand{\affilref}[1]{%
  \@ifundefined{c@#1}%
    {\newcounter{#1}%
     \setcounter{#1}{\theaffilct}%
     \refstepcounter{affilct}%
     \label{#1}%
     }{}%
  \ref{#1}%
 }
\makeatother

\makeatletter
\newcommand*\affilreftxt[2]{%
  \@ifundefined{c@#1txt}
    {\newcounter{#1txt}%
     \setcounter{#1txt}{1}
     \altaffiltext{\ref{#1}}{#2}
     }{
     }
  }
\makeatother


\author{Adam G. Riess\altaffilmark{\affilref{JHU},\affilref{STScI}}}
\affilreftxt{JHU}{\JHU}
\affilreftxt{STScI}{\STScI}
\email{ariess@stsci.edu}

\author{Steven A. Rodney\altaffilmark{\affilref{USC}}}
\affilreftxt{USC}{\USC}

\author{Daniel M. Scolnic\altaffilmark{\affilref{KICP}}}
\affilreftxt{KICP}{\KICP}

\author{Daniel L. Shafer\altaffilmark{\affilref{JHU}}}
\affilreftxt{JHU}{\JHU}

\author{Louis-Gregory Strolger\altaffilmark{\affilref{STScI}}}
\affilreftxt{STScI}{\STScI}

\author{Henry C. Ferguson\altaffilmark{\affilref{STScI}}}
\affilreftxt{STScI}{\STScI}

\author{Marc Postman\altaffilmark{\affilref{STScI}}}
\affilreftxt{STScI}{\STScI}

\author{Or Graur\altaffilmark{\affilref{CfA},\affilref{AMNH},\affilref{NSF}}}
\affilreftxt{CfA}{\CfA}
\affilreftxt{AMNH}{\AMNH}
\affilreftxt{NSF}{\NSF}

\author{Dan Maoz\altaffilmark{\affilref{TelAviv}}}
\affilreftxt{TelAviv}{\TelAviv}

\author{Saurabh W. Jha\altaffilmark{\affilref{Rutgers}}}
\affilreftxt{Rutgers}{\Rutgers}

\author{Bahram Mobasher\altaffilmark{\affilref{Riverside}}}
\affilreftxt{Riverside}{\Riverside}

\author{Stefano Casertano\altaffilmark{\affilref{STScI}}}
\affilreftxt{STScI}{\STScI}

\author{Brian Hayden\altaffilmark{\affilref{LBNL},\affilref{BerkeleyPhys}}}
\affilreftxt{LBNL}{\LBNL}
\affilreftxt{BerkeleyPhys}{\BerkeleyPhys}

\author{Alberto Molino\altaffilmark{\affilref{SaoPaulo}}}
\affilreftxt{SaoPaulo}{\SaoPaulo}

\author{Jens Hjorth\altaffilmark{\affilref{DARK}}}
\affilreftxt{DARK}{\DARK}

\author{Peter M. Garnavich\altaffilmark{\affilref{NotreDame}}}
\affilreftxt{NotreDame}{\NotreDame}

\author{David O. Jones\altaffilmark{\affilref{JHU}}}
\affilreftxt{JHU}{\JHU}

\author{Robert P. Kirshner\altaffilmark{\affilref{CfA}}}
\affilreftxt{CfA}{\CfA}

\author{Anton M. Koekemoer\altaffilmark{\affilref{STScI}}}
\affilreftxt{STScI}{\STScI}

\author{Norman A. Grogin\altaffilmark{\affilref{STScI}}}
\affilreftxt{STScI}{\STScI}

\author{Gabriel Brammer\altaffilmark{\affilref{STScI}}}
\affilreftxt{STScI}{\STScI}

\author{Shoubaneh Hemmati\altaffilmark{\affilref{IPAC}}}
\affilreftxt{IPAC}{\IPAC}

\author{Mark Dickinson\altaffilmark{\affilref{NOAO}}}
\affilreftxt{NOAO}{\NOAO}

\author{Peter M. Challis\altaffilmark{\affilref{CfA}}}
\affilreftxt{CfA}{\CfA}

\author{Schuyler Wolff\altaffilmark{\affilref{JHU}}}
\affilreftxt{JHU}{\JHU}

\author{Kelsey I. Clubb\altaffilmark{\affilref{Berkeley}}}
\affilreftxt{Berkeley}{\Berkeley}

\author{Alexei V. Filippenko\altaffilmark{\affilref{Berkeley},\affilref{MillerFellow}}}
\affilreftxt{Berkeley}{\Berkeley}
\affilreftxt{MillerFellow}{\MillerFellow}

\author{Hooshang Nayyeri\altaffilmark{\affilref{Irvine}}}
\affilreftxt{Irvine}{\Irvine}

\author{Vivian U\altaffilmark{\affilref{Riverside},\affilref{Irvine},\affilref{ChancellorFellow}}}
\affilreftxt{Riverside}{\Riverside}
\affilreftxt{Irvine}{\Irvine}
\affilreftxt{ChancellorFellow}{\ChancellorFellow}

\author{David C. Koo\altaffilmark{\affilref{UCSC}}}
\affilreftxt{UCSC}{\UCSC}

\author{Sandra M. Faber\altaffilmark{\affilref{UCSC}}}
\affilreftxt{UCSC}{\UCSC}

\author{Dale Kocevski\altaffilmark{\affilref{Colby}}}
\affilreftxt{Colby}{\Colby}

\author{Larry Bradley\altaffilmark{\affilref{STScI}}}
\affilreftxt{STScI}{\STScI}

\author{Dan Coe\altaffilmark{\affilref{STScI}}}
\affilreftxt{STScI}{\STScI}

\begin{abstract}

We present an analysis of 15 Type Ia supernovae (SNe~Ia) at redshift $z > 1$ (9 at $1.5 < z < 2.3$) recently discovered in the CANDELS and CLASH Multi-Cycle Treasury programs using WFC3 on the \textit{Hubble Space Telescope}. We combine these SNe~Ia with a new compilation of $\sim\!1050$ SNe~Ia, jointly calibrated and corrected for simulated survey biases to produce accurate distance measurements. We present unbiased constraints on the expansion rate at six redshifts in the range $0.07 < z < 1.5$ based only on this combined SN~Ia sample. The added leverage of our new sample at $z > 1.5$ leads to a factor of $\sim\!3$ improvement in the determination of the expansion rate at $z = 1.5$, reducing its uncertainty to $\sim\!20$\%, a measurement of $H(z = 1.5)/H_0 = 2.67^{+0.83}_{-0.52}$. We then demonstrate that these six measurements alone provide a nearly identical characterization of dark energy as the full SN sample, making them an efficient compression of the SN~Ia data. The new sample of SNe~Ia at $z > 1.5$ usefully distinguishes between alternative cosmological models and unmodeled evolution of the SN~Ia distance indicators, placing empirical limits on the latter. Finally, employing a realistic simulation of a potential \textit{WFIRST} SN survey observing strategy, we forecast optimistic future constraints on the expansion rate from SNe~Ia.

\end{abstract}


\section{Introduction} \label{sec:Introduction}

Type Ia supernovae (SNe~Ia) at redshift $z > 1$ offer unique leverage on investigations relating to the nature of their progenitors, their accuracy as distance indicators, and the parameters of the cosmological model. Unfortunately, ground-based facilities are extremely challenged to produce reliable discoveries of SNe~Ia at $z > 1$, a task demanding significant and repeatable detections and robust classifications at $I \sim 26~\text{mag}$.

Thus, for the past two decades, the \textit{Hubble Space Telescope} (\textit{HST}) has offered the best perch from which to harvest these objects, with the rate of collection limited only by its relatively modest field of view. The first robust, multi-object sample of SNe~Ia at $z > 1$ came from searching the GOODS fields with the \textit{HST} Advanced Camera for Surveys (ACS) and its $z$-band filter, with crucial near-infrared follow-up observations of the rest-frame optical light obtained using NICMOS and confirming spectroscopy from the ACS grism. The first sample of 7 SNe~Ia at $z > 1.25$ provided a crucial check that dimming from astrophysical effects was not mimicking cosmic acceleration \citep{Riess:2004b}. A follow-up program increased the sample of reliable SNe~Ia at $z > 1$ to 18 \citep{Riess:2007} followed by another 12 from targeting cluster fields \citep{Suzuki:2012}. This sample of $\sim\!30$ successfully extended the SN~Ia measurement of expansion to the matter-dominated era to break degeneracies between dark energy and dark matter.

Still, clues available only at $z > 1.5$ beckoned. Owing to the red-limit of \textit{HST} CCDs and the roughly Gyr delay between progenitor formation and SN~Ia explosion \citep{Rodney:2014}, only $\sim\!3$ moderately constrained SNe~Ia at $z > 1.5$ were previously discovered with \textit{HST}: SN 1997ff at $z = 1.755$, SN 2003ak at $z = 1.551$, and SCP0401 at $z = 1.713$ \citep{Gilliland:1999,Riess:2001,Riess:2004b,Rubin:2013}. An effective program to find SNe~Ia at $z > 1.5$ required WFC3-IR, the first wide-area (greater than an arcminute) infrared HgCdTe detector on \textit{HST}, installed in 2009, which extended the red cutoff to 1.6~$\mu$m. Two of the initial three Multi-Cycle Treasury (MCT) programs with WFC3, CANDELS \citep[PI: Faber and Ferguson,][]{Grogin:2011,Koekemoer:2011} and CLASH \citep[PI: Postman,][]{Postman:2012} were selected to enable the discovery of SNe~Ia at $z > 1.5$ with an additional program of coordinated SN follow-up observations \citep[PI: Riess,][]{Rodney:2014,Graur:2014}. These MCT programs were three-year extragalactic imaging campaigns initiated in \textit{HST} Cycle 18, beginning October 2010. Both MCT programs employed ACS and WFC3-IR with cadences of $\sim\!50$ days between epochs, chosen to match the risetime of SNe~Ia time-dilated to $1.5 < z < 2.0$. \citet[][hereafter \R14]{Rodney:2014} comprehensively described the SN search component of the CANDELS program and measured the volumetric SN~Ia rate from the complete CANDELS sample of 65~SNe out to $z = 2.5$. \citet[][hereafter \G14]{Graur:2014} presented the SN~Ia rates analysis from the CLASH program, using a sample of 27~SNe detected in the \textit{HST} parallel fields ($\sim\!6$\arcmin\ from the galaxy clusters that make up the primary targets for CLASH). For full details of the survey design and observations, we refer the reader to \R14 and \G14.

These programs together identified 15~SNe~Ia at $z > 1$, 9 of which (7 at $z > 1.5$) are sufficiently well-measured to derive reliable distance estimates. Detailed studies of the first two such events were presented by \cite{Rodney:2012} and \cite{Jones:2013}, and a novel approach to SN classification via medium-band infrared imaging was presented for two others by \cite{Rodney:2015b}.

Here for the first time we derive a set of distance estimates for this sample calibrated for a joint cosmological analysis with a compilation of SNe~Ia from previous surveys \citep[][in prep]{Scolnic:2017}. The most significant augmentation of the extant SN~Ia sample is the set of SN~Ia distances presented here at $z > 1.5$, which usefully extends the SN-based determination of the expansion rate of the universe to a higher redshift, $z \approx 1.5$, than previously possible. In Section~\ref{sec:TheSample}, we present details of the SN sample, and in Section~\ref{sec:Hz}, we present constraints on the scale-free expansion history and carry out some related investigations. We summarize our conclusions in Section~\ref{sec:conclude}.

\section{SN Ia Sample} \label{sec:TheSample}

\begin{deluxetable*}{llllll} 
\tablecolumns{10}
\tablecaption{SNe Ia from CANDELS + CLASH at $z > 1$ \label{tab:Coordinates}}
\tablehead{ 
  \colhead{SN ID}
  & \colhead{Nickname}
  & \colhead{Survey}
  & \colhead{Field}
  & \colhead{$\alpha$(J2000)}
  & \colhead{$\delta$(J2000)}
}
\startdata
CLA10Cal & Caligula   & CLASH   & Abell 383 IR par & 02:48:25.74 & $-$03:33:08.8 \\ 
CLF11Ves & Vespasian  & CLASH   & MACS2129 ACS par & 21:29:42.60 & $-$07:41:47.7 \\ 
CLH11Tra & Trajan     & CLASH   & MS2137 ACS par   & 21:39:46.05 & $-$23:38:34.8 \\ 
CLP12Get & Geta       & CLASH   & RXJ2129 IR par   & 21:29:23.89 & $+$00:08:24.8 \\ 
COS12Car & Carter     & CANDELS & COSMOS           & 10:00:14.72 & $+$02:11:32.6 \\ 
EGS11Oba & Obama      & CANDELS & EGS              & 14:20:32.66 & $+$53:02:48.2 \\ 
EGS13Rut & Rutledge   & CANDELS & EGS              & 14:20:48.11 & $+$53:04:22.1 \\ 
GND12Col & Colfax     & CANDELS & GOODS-N Deep     & 12:36:37.58 & $+$62:18:33.1 \\ 
GND13Cam & Camille    & CANDELS & GOODS-N Deep     & 12:37:07.37 & $+$62:10:26.9 \\ 
GND13Gar & Garner     & CANDELS & GOODS-N Deep     & 12:36:40.81 & $+$62:11:14.2 \\ 
GND13Jay & Jay        & CANDELS & GOODS-N Deep     & 12:36:41.38 & $+$62:11:30.1 \\ 
GND13Sto & Stone      & CANDELS & GOODS-N Deep     & 12:37:16.77 & $+$62:16:41.4 \\ 
GSD10Pri & Primo      & CANDELS & GOODS-S Deep     & 03:32:38.01 & $-$27:46:39.1 \\ 
GSD11Was & Washington & CANDELS & GOODS-S Deep     & 03:32:20.85 & $-$27:49:41.5 \\ 
UDS10Wil & Wilson     & CANDELS & UDS              & 02:17:46.33 & $-$05:15:24.0 \\ 
\enddata
\end{deluxetable*}

\begin{deluxetable*}{lllllll}
\tablecolumns{10}
\tablecaption{Final Redshifts and Classifications \label{tab:RedshiftAndClassification}}
\tablehead{\colhead{SN ID} & \colhead{Redshift\tablenotemark{a}} & \colhead{Redshift Source\tablenotemark{b}} & \colhead{$P$(Ia)\tablenotemark{c}} & \colhead{Supporting Evidence\tablenotemark{d}} & \colhead{Confidence\tablenotemark{e}} & \colhead{Primary Reference\tablenotemark{f}}
}
\startdata
CLA10Cal & 1.800$\pm$0.1                   & phot-$z$                                   & 0.95     & \nodata               & bronze & \citet{Graur:2014} \\
CLF11Ves & 1.206$\pm$0.007                 & spec-$z$ (\textit{HST}+G800L)              & $>$0.99  & spec, early-type host & gold   & \citet{Graur:2014} \\
CLH11Tra & 1.520$\pm$0.04\tablenotemark{g} & phot-$z$                                   & $>$0.99  & early-type host       & gold   & \citet{Graur:2014} \\
CLP12Get & 1.700$\pm$0.04                  & phot-$z$                                   & $>$0.99  & early-type host       & gold   & \citet{Graur:2014} \\
COS12Car & 1.540$\pm$0.04                  & SN spec-$z$ (\textit{HST}+G141)            & $>$0.99  & spec                  & gold   & \citet{Rodney:2014} \\
EGS11Oba & 1.409$\pm$0.002                 & spec-$z$ (Keck+LRIS,DEIMOS)                & 0.9      & \nodata               & bronze & \citet{Rodney:2014} \\
EGS13Rut & 1.614$\pm$0.005                 & spec-$z$ (\textit{HST}+G141, single line)  & $>$0.99  & \nodata               & silver & \citet{Rodney:2014} \\
GND12Col & 2.260$^{+0.02}_{-0.10}$         & phot-$z$                                   & $>$0.99  & med. band             & gold   & \citet{Rodney:2015b} \\
GND13Cam & 1.222$\pm$0.002                 & spec-$z$ (AGHAST, \textit{HST}+G141)       & $>$0.99  & \nodata               & silver & \citet{Rodney:2014} \\
GND13Gar & 1.070$\pm$0.02                  & SN spec-$z$ (\textit{HST}+G800L)           & $>$0.99  & spec                  & gold   & \citet{Rodney:2014} \\
GND13Jay & 1.030$\pm$0.01                  & spec-$z$ (AGHAST, \textit{HST}+G141)       & $>$0.99  & \nodata               & silver & \citet{Rodney:2014} \\
GND13Sto & 1.800$\pm$0.02                  & spec-$z$                                   & $>$0.99  & med. band             & gold   & \citet{Rodney:2015b} \\
GSD10Pri & 1.550$\pm$0.0001                & spec-$z$                                   & $>$0.99  & spec                  & gold   & \citet{Rodney:2012} \\
GSD11Was & 1.330$\pm$0.02                  & spec-$z$ (\textit{HST}+G141)               & $>$0.99  & spec                  & gold   & \citet{Rodney:2014} \\
UDS10Wil & 1.914$\pm$0.001                 & spec-$z$                                   & $>$0.99  & spec                  & gold   & \citet{Jones:2013} \\
\enddata
\tablenotetext{a}{Final composite redshift, incorporating all evidence from SN and host.}
\tablenotetext{b}{All phot-$z$ and spec-$z$ redshifts are principally constrained by the host galaxy, except where a SN spec-$z$ is noted.}
\tablenotetext{c}{Classification probability from the SN light curve, including host redshift priors, using STARDUST (\R14).}
\tablenotetext{d}{Additional factors influencing the classification confidence. ``spec'': SN spectrum; ``med. band'': pseudocolors from medium-band infrared imaging; ``early-type host'': host galaxy is identified as an early-type galaxy, unlikely to host core-collapse SNe.}
\tablenotetext{e}{Confidence in the Type Ia SN classification.}
\tablenotetext{f}{Primary reference for further information on discovery, redshift, and classification.}
\tablenotetext{g}{Revised from \citet{Graur:2014}}
\end{deluxetable*}

\begin{deluxetable*}{lllllll} 
\tablecolumns{10}
\tablecaption{SN Host Galaxy Data \label{tab:HostGalaxies}}
\tablehead{ 
  \colhead{SN ID}
  & \colhead{Host $\alpha$(J2000)}
  & \colhead{Host $\delta$(J2000)}
  & \colhead{Host Redshift\tablenotemark{a}}
  & \colhead{Morphology}
  & \colhead{Star Formation}
  & \colhead{Redshift Source}
}
\startdata
CLA10Cal & 02:48:25.74    & $-$03:33:08.8 & 1.8$\pm$0.1           & spheroid/disk    & active       &  phot-$z$      \\   
CLF11Ves & 21:29:42.62    & $-$07:41:47.5 & 1.206$\pm$0.007       & spheroid         & passive      &  \textit{HST}+ACS     \\   
CLH11Tra & 21:39:46.04    & $-$23:38:34.6 & 1.52$\pm$0.04         & spheroid         & passive      &  phot-$z$      \\   
CLP12Get & 21:29:23.92    & $+$00:08:23.8 & 1.70$\pm$0.04         & spheroid         & passive      &  phot-$z$                       \\   
COS12Car\tablenotemark{b} &   10:00:14.72 & $+$02:11:32.6 &    \nodata            & undetected       & undetected   &  \nodata                       \\   
EGS11Oba & 14:20:32.67    & $+$53:02:48.1 & 1.409$\pm$0.002       & disk/irregular   & active       &  Keck+LRIS      \\   
EGS13Rut & 14:20:48.11    & $+$53:04:22.1 & 1.614$\pm$0.005       & disk             & active       &  \textit{HST}+WFC3            \\  
GND12Col & 12:36:37.51    & $+$62:18:32.6 & 2.260$^{+0.02}_{-0.10}$  & spheroid        & active       & phot-$z$               \\  
GND13Cam & 12:37:07.38    & $+$62:10:27.2 & 1.222$\pm$0.002       & spheroid/disk    & starburst    &  \textit{HST}+WFC3           \\  
GND13Gar & 12:36:40.80    & $+$62:11:14.6 & 1.86$\pm$0.77         & undefined        & starburst    &  phot-$z$             \\  
GND13Jay & 12:36:41.37    & $+$62:11:29.5 & 1.03$\pm$0.01         & disk             & active       &  \textit{HST}+WFC3           \\  
GND13Sto & 02:37:16.59    & $+$62:16:43.4 & 1.80$\pm$0.02         & undefined        & active       &  phot-$z$             \\  
GSD10Pri & 03:32:37.99    & $-$27:46:38.7 & 1.550$\pm$0.0001      & irregular        & starburst    &  VLT+X-Shooter                 \\   
GSD11Was & 03:32:20.86    & $-$27:49:41.5 & 1.042$\pm$0.23        & disk             & starburst    &  \textit{HST}+WFC3            \\   
UDS10Wil & 02:17:46.33    & $-$05:15:23.9 & 1.914$\pm$0.001       & spheroid         & starburst    &  VLT+X-Shooter                 \\   
\enddata
\tablenotetext{a}{Photometric redshifts are marked as ``phot-$z$'' and spectroscopic redshifts are labeled with the observatory and instrument employed.}
\tablenotetext{b}{No plausible host galaxy was identified for SN COS12Car.  The coordinates given are for the SN itself.}
\end{deluxetable*}

\begin{deluxetable*}{lcccccc}
\tablecaption{SALT2 Light-Curve Fit Parameters \label{tab:distances}}
\tablehead{\colhead{SN ID} & \colhead{$m_B$} & \colhead{$x_1$} & \colhead{$c$} & \colhead{$\Delta\mu_\text{bias-corr}$} & \colhead{$\mu$\,(mag)} & \colhead{Notes}}
\startdata
CLA10Cal &                -- &               -- &                 -- &      -- &               -- & poor light-curve fit \\
CLF11Ves & $25.38$ ($0.091$) & $-1.24$ ($0.60$) & $-0.288$ ($0.101$) & $+0.27$ & $25.73$ ($0.34$) & \\
CLH11Tra & $25.30$ ($0.095$) & $-3.35$ ($2.10$) & $-0.272$ ($0.090$) &      -- &               -- & fails $x_1$ cut ($x_1 < -3$, $\sigma_{x_1} > 1$) \\
CLP12Get & $25.73$ ($0.088$) & $+1.01$ ($0.95$) & $-0.139$ ($0.098$) & $+0.18$ & $26.06$ ($0.28$) & \\
COS12Car & $26.14$ ($0.122$) & $+2.35$ ($0.83$) & $+0.152$ ($0.083$) & $+0.07$ & $25.91$ ($0.21$) & \\
EGS11Oba &                -- &               -- &                 -- &      -- &               -- & poor light-curve fit \\
EGS13Rut & $25.92$ ($0.071$) & $+0.98$ ($1.08$) & $+0.055$ ($0.046$) & $-0.07$ & $25.93$ ($0.20$) & \\
GND12Col & $26.81$ ($0.056$) & $+0.02$ ($0.91$) & $+0.128$ ($0.133$) & $-0.50$ & $26.88$ ($0.25$) & \\
GND13Cam & $25.91$ ($0.061$) & $-1.35$ ($0.48$) & $-0.083$ ($0.043$) &      -- &               -- & Hubble diagram outlier ($> 4\sigma$) \\
GND13Gar & $25.42$ ($0.259$) & $+0.02$ ($0.99$) & $+0.310$ ($0.179$) &      -- &               -- & fails color cut ($c > 0.3$) \\
GND13Jay & $24.56$ ($0.672$) & $-2.04$ ($0.92$) & $-0.373$ ($0.447$) &      -- &               -- & fails color cut ($c < -0.3$) \\
GND13Sto & $26.15$ ($0.074$) & $-0.48$ ($0.70$) & $+0.000$ ($0.071$) & $-0.17$ & $26.20$ ($0.19$) & \\
GSD10Pri & $25.76$ ($0.089$) & $-0.51$ ($0.41$) & $-0.186$ ($0.078$) & $+0.16$ & $26.01$ ($0.19$) & \\
GSD11Was & $25.32$ ($0.057$) & $+1.04$ ($0.67$) & $-0.089$ ($0.039$) & $+0.09$ & $25.60$ ($0.15$) & \\
UDS10Wil & $26.28$ ($0.172$) & $-1.64$ ($0.76$) & $+0.082$ ($0.152$) & $-0.43$ & $26.15$ ($0.26$) & \\
\enddata
\end{deluxetable*}

\begin{deluxetable*}{llllll}
\tablecolumns{10}
\tablecaption{SNe Ia at $z > 1$ from Other Surveys \label{tab:OtherSurveys}}
\tablehead{ 
  \colhead{SN ID}
  & \colhead{Nickname}
  & \colhead{Survey\tablenotemark{a}}
  & \colhead{Confidence\tablenotemark{b}}
  & \colhead{Redshift}
  & \colhead{Reference}
}
\startdata
1997ff      & 1997ff      & HDFN           & Gold    & 1.755  & \citealt{Riess:2001}   \\
2002fw      & Aphrodite   & Higher-z GOODS & Gold    & 1.30   & \citealt{Riess:2004b}   \\
2002fx      & Athena      & Higher-z GOODS & Silver  & 1.40   & \citealt{Riess:2004b}   \\
2002hp      & Thoth       & Higher-z GOODS & Gold    & 1.305  & \citealt{Riess:2004b}   \\
2002ki      & Nanna       & Higher-z GOODS & Gold    & 1.141  & \citealt{Riess:2004b}   \\
2003aj      & Inanna      & Higher-z GOODS & Silver  & 1.307  & \citealt{Riess:2004b}   \\
2003ak      & Gilgamesh   & Higher-z GOODS & Silver  & 1.551  & \citealt{Riess:2004b}   \\
2003az      & Torngasak   & Higher-z GOODS & Silver  & 1.265  & \citealt{Riess:2004b}   \\
2003dy      & Borg        & Higher-z GOODS & Gold    & 1.34   & \citealt{Riess:2004b}   \\
HST04Eag    & Eagle       & Higher-z PANS & Gold    & 1.019  & \citealt{Riess:2007}   \\
HST04Gre    & Greenburg   & Higher-z PANS & Gold    & 1.14   & \citealt{Riess:2007}   \\
HST04Mcg    & Mcguire     & Higher-z PANS & Gold    & 1.357  & \citealt{Riess:2007}   \\
HST04Sas    & Sasquatch   & Higher-z PANS & Gold    & 1.39   & \citealt{Riess:2007}   \\
HST05Fer    & Ferguson    & Higher-z PANS & Gold    & 1.02   & \citealt{Riess:2007}   \\
HST05Gab    & Gabi        & Higher-z PANS & Gold    & 1.12   & \citealt{Riess:2007}   \\
HST05Koe    & Koekemoer   & Higher-z PANS & Gold    & 1.23   & \citealt{Riess:2007}   \\
HST05Lan    & Lancaster   & Higher-z PANS & Gold    & 1.235  & \citealt{Riess:2007}   \\
HST05Str    & Strolger    & Higher-z PANS & Gold    & 1.027  & \citealt{Riess:2007}   \\
SCP0401     & SCP0401     & SCP GOODS & Gold    & 1.713  & \citealt{Rubin:2013}  \\
SCP05D0     & Frida       & SCP CSS  & Gold    & 1.014  & \citealt{Suzuki:2012}  \\
SCP05D6     & Maggie      & SCP CSS  & Gold    & 1.315  & \citealt{Suzuki:2012}  \\
SCP06A4     & Aki         & SCP CSS  & Silver  & 1.192  & \citealt{Suzuki:2012}  \\
SCP06C0     & Noa         & SCP CSS  & Gold    & 1.092  & \citealt{Suzuki:2012}  \\
SCP06F12    & Caleb       & SCP CSS  & Silver  & 1.110  & \citealt{Suzuki:2012}  \\
SCP06G4     & Shaya       & SCP CSS  & Gold    & 1.35   & \citealt{Suzuki:2012}  \\
SCP06H5     & Emma        & SCP CSS  & Gold    & 1.231  & \citealt{Suzuki:2012}  \\
SCP06K0     & Tomo        & SCP CSS  & Gold    & 1.415  & \citealt{Suzuki:2012}  \\
SCP06K18    & Alexander   & SCP CSS  & Silver  & 1.411  & \citealt{Suzuki:2012}  \\
SCP06N33    & Naima       & SCP CSS  & Silver  & 1.188  & \citealt{Suzuki:2012}  \\
SCP06R12    & Jennie      & SCP CSS  & Gold    & 1.212  & \citealt{Suzuki:2012}  \\
SCP06U4     & Julia       & SCP CSS  & Gold    & 1.05   & \citealt{Suzuki:2012}  \\
\enddata
\tablenotetext{a}{
\textit{HDFN:} SN 1997ff was discovered in observations of the Hubble
Deep Field North \citep[HDFN;][]{Gilliland:1999,Dickinson:2001}.\\
\textit{Higher-z GOODS/PANS:} Discoveries by the Hubble Higher-z SN
Search team, from the SN component of the Great Observatories Origins
Deep Survey \citep[GOODS, HST-GO-9728, HST-GO-9352,
  HST-GO-9583;][]{Giavalisco:2004,Strolger:2004} and the successor
program Probing Acceleration Now with Supernova \citep[PANS,
  HST-GO-10339;][]{Riess:2007}.\\
\textit{SCP-GOODS/CSS:} Discoveries by the Supernova Cosmology Project
(SCP) on the GOODS fields (HST-GO-9727) or in the Cluster Supernova
Search \citep[CSS, HST-GO-9425;][]{Dawson:2009}.
}
\tablenotetext{b}{Confidence in the Type Ia classification, as reported by \cite{Riess:2007} or \cite{Suzuki:2012}, where the latter have been translated from ``secure/probable/plausible'' to ``gold/silver/bronze.''}
\end{deluxetable*}

From the total set of 92 CANDELS and CLASH SNe, we have identified 15 as likely SNe~Ia at $z > 1$ with sufficient confidence for use as distance indicators. We present the coordinates of these objects in Table~\ref{tab:Coordinates}, their redshifts and classifications in Table~\ref{tab:RedshiftAndClassification}, the properties of their host galaxies in Table~\ref{tab:HostGalaxies}, and their distance-related parameters in Table~\ref{tab:distances}. For inclusion in this subset, we require at least enough samplings of the light and color curves to exceed the number of free parameters in the light-curve fit. This effectively means that we require a minimum of four independent observation epochs, providing at least a modicum of constraint on the light-curve shape. We also require that the first epoch with $>\!3\sigma$ detection must be no more than 10~days after the peak of the light curve in the rest-frame $B$ band, consistent with the requirements used by \cite{Riess:1996} and \cite{Riess:2007}. Finally, we require that at least one of the epochs includes WFC3-IR observations in both the $F125W$ and $F160W$ bands, which provide a measurement of the SN color at rest-frame optical wavelengths for $1 < z < 2.5$.

For the cosmological analysis presented here, we further subdivide this sample into three confidence categories: \textit{gold}, \textit{silver}, and \textit{bronze}, following the convention of \cite{Strolger:2004}, \cite{Riess:2004b}, and \cite{Riess:2007}. The gold sample comprises those SNe with compelling classifications as Type Ia, while the silver label indicates a ``very likely'' Type Ia classification, and the bronze objects are those that are probably Type Ia, but have some nonnegligible probability of misclassification.

As detailed by \R14\ and \G14, the classifications of these SNe at $z > 1$ sometimes rely on photometric evidence. Spectra are available for 6 of the 15 SNe~Ia at $z > 1$ (3 at $z > 1.5$), while two others use medium bands to measure the strength of SN~Ia spectral features. This mixture of classification methods is necessitated by the difficulty of achieving a purely spectroscopic classification for such high-redshift SNe \citep[see, e.g.,][]{Rodney:2012,Frederiksen:2012px,Rubin:2013,Jones:2013}. Photometric classification of these SNe was performed using STARDUST\footnote{STARDUST: Supernova Taxonomy And Redshift Determination Using SNANA Templates}, a Bayesian algorithm employing a comparison of multi-band light curves against 43 template-based models representing Type Ia and core-collapse SNe (\R14). For inclusion in the gold and silver samples, we require a Type Ia classification probability $P(\text{Ia}) > 0.99$; the two objects with $0.9 < P(\text{Ia}) < 0.99$ were relegated to the bronze sample.

The gold objects are further distinguished by having at least one piece of corroborating evidence to support the Type Ia classification. For 6 objects, we have a spectroscopic observation that is well-matched by a SN~Ia spectral template, presented by \R14\ and \G14. Two more SNe have medium-band infrared imaging that provides evidence for Type Ia \textit{spectral features} in medium-band minus broad-band pseudocolors \citep{Rodney:2015b}. Finally, three of the gold sample SNe have a host galaxy that is classified as ``early type'' based on morphology and colors, indicating an old stellar population that would be unlikely to host a core-collapse SN \citep{Riess:2001}.

We discard the 2 bronze SNe whose classification is too uncertain and proceed with the analysis of the remaining 13 gold and silver SNe at $z > 1$ (8 at $z > 1.5$) from the CANDELS and CLASH programs. Assuming the 3 silver SNe in the sample are Type Ia with 99\% confidence, there is a $\sim\!97$\% chance that all of the SNe in the cosmological analysis are Type Ia. We combine this set (hereafter, the MCT set) with a uniformly calibrated compilation of $\sim\!1050$ spectroscopically classified SNe~Ia, the Pantheon compilation \citep{Scolnic:2017}. This compilation includes SNe from the Harvard-Smithsonian Center for Astrophysics SN surveys \citep[CfA,][]{Hicken:2009a}, the Carnegie Supernova Project \citep[CSP,][]{Stritzinger:2011}, the Sloan Digital Sky Survey \citep[SDSS,][]{Kessler:2009b}, the Pan-STARRS1 Medium-Deep Survey \citep[PS1,][]{Rest:2014}, and the Canada-France-Hawaii Telescope Supernova Legacy Survey \citep[SNLS,][]{Conley:2011}. The compilation includes all SNe from the \cite{Rest:2014} sample and from the samples included in the joint light-curve analysis \citep[JLA;][]{Betoule:2014}, all uniformly calibrated as presented in the Supercal analysis \citep{Scolnic:2015eyc}. The Pantheon compilation also includes 12 equivalently high-confidence SNe~Ia at $1 < z < 1.4$ from past \textit{HST} SN surveys (see Table~\ref{tab:OtherSurveys}), 9 from \cite{Riess:2004b} and \cite{Riess:2007} and 3 from \cite{Suzuki:2012}, that meet the criteria given in \cite{Scolnic:2017}.

\section{High-Redshift Measurements of the Hubble Parameter} \label{sec:Hz}

At $z \gtrsim 1$, dark energy is a small contribution to the energy budget ($\rho_\Lambda/\rho \approx 0.2$ at $z = 1$ and $\approx\!0.1$ at $z = 1.5$) and therefore has a small effect on dynamics. With abundant and better-measured SNe at lower redshifts, constraints on typical one-or-two-parameter dark energy models are only weakly improved by observations of SNe at $z > 1$ \citep[see also][regarding $z > 2$]{Andersen:2017uew}. This is especially true for combined constraints when precise distances from cosmic microwave background (CMB) and baryon acoustic oscillations (BAO) measurements are included.

Nevertheless, the new SNe at $z > 1.5$ presented here allow us to constrain the (dimensionless) Hubble parameter $E(z) \equiv H(z)/H_0$ at greater redshifts than previously possible. The quantity $H(z)$ is particularly useful because it is both a direct probe of cosmology and still closely tied to the data. As a dynamical quantity, $H(z)$ contains information about the expansion history without reference to any physical cosmological model. Also, at least for current SN~Ia data, the inferred $H(z)$ measurements are fairly local; that is, they are predominantly influenced by SNe at nearby redshifts. The quantity $E(z)$, which contains similarly useful information but can be measured using SN~Ia data alone, makes the results independent of uncertainties associated with the determination of the absolute distance scale of SNe~Ia \citep{Riess:2016jrr}.

As a direct probe of the expansion rate ($H \equiv \dot{a}/a$), measurements of $E(z)$ are particularly dense with cosmological information. They provide, for instance, a straightforward way to test or falsify a given cosmological model \citep{Mortonson:2008qy,Mortonson:2009hk,Shafieloo:2009hi}. Given current constraints on its parameters, the flat $\Lambda$CDM model already makes very precise predictions for such basic observables. Constraints on the matter density $\Omega_m$ from combined probes \citep[e.g.,][]{Ade:2015xua} imply that $E(z)$, defined to be exactly one at $z = 0$, is predicted to a precision ranging from $\sim\!0.1$\% at $z = 0.1$ to $\sim\!1$\% at $z = 2$. Therefore, any new, independent measurement of $E(z)$, particularly in a new redshift range, is a direct and nontrivial test of the standard cosmological model. Given the present $>\!3\sigma$ tension between $H(z)$ calibrated at $z \approx 0$ \citep{Riess:2016jrr} and at $z \approx 1100$ by the CMB \citep{Ade:2015xua}, it is especially worthwhile to see if the expansion rate fails to match the standard $\Lambda$CDM model prediction anywhere along this redshift range.

Furthermore, as we will illustrate, accurate estimates of $E(z)$ from SN~Ia data are a convenient and efficient form of data compression, allowing one to obtain SN~Ia constraints on dark energy and other cosmological parameters quickly and robustly using a very small and easily provided set of measurements. Such data compression techniques will be especially useful as SN~Ia samples grow significantly in size in the coming decade. Some recent SN~Ia analyses \citep[e.g.,][]{Betoule:2014} have included compressed versions of the SN data in the form of binned distance moduli, and it is worth investigating the extent to which $E(z)$ measurements can serve a similar purpose.

Finally, quantifying SN~Ia constraints on $E(z)$ facilitates a more direct comparison with other cosmological probes of geometry, such as anisotropic fits of the BAO feature, which effectively constrain a dimensionless measure of the expansion rate, the product of the Hubble parameter and the sound horizon, where the latter is inferred precisely from CMB observations.

Our aim here is to employ a new, well-calibrated compilation of SNe~Ia, featuring the final addition of 9 new SNe~Ia at $z > 1$ from the CANDELS and CLASH programs, to obtain unbiased estimates of the Hubble parameter $E(z)$ up to $z \approx 1.5$.

In what follows, we will briefly review some proposed methodologies for inferring $E(z)$ from SN~Ia data (Section~\ref{sec:methods}) and then discuss our approach and how it overcomes some important limitations (Section~\ref{sec:interpfits}). In Section~\ref{sec:results}, we present constraints on $E(z)$ for the Pantheon SN compilation supplemented by the MCT SNe (i.e., Pantheon + MCT\footnote{Note that the Pantheon compilation as defined in \cite{Scolnic:2017} includes the MCT SNe presented here.}). We illustrate how the handful of high-redshift SNe from CANDELS and CLASH significantly improves the determination of $E(z)$ at $z \approx 1.5$. We also illustrate the effectiveness of the $E(z)$ measurements in subsequent inference of cosmological parameters, and, in Section~\ref{sec:SN_evol}, the ability of the high-redshift SNe~Ia to distinguish cosmology from SN~Ia evolution. Finally, employing a realistic simulation of a potential \textit{WFIRST} SN survey observing strategy, we compare our current results with optimistic future constraints on $E(z)$ (Section~\ref{sec:wfirst}).

\subsection{SN Ia Measurements of \texorpdfstring{$E(z)$}{E(z)}} \label{sec:methods}

SNe~Ia measure distances most directly; roughly speaking, each SN provides an independent measurement of the luminosity distance to its redshift. For a flat universe, we have
\begin{equation}
d_L(z) = \frac{c}{H_0} (1 + z) \int_0^{z} \frac{dz'}{E(z')} \ ;
\end{equation}
therefore the (inverse) Hubble parameter, the derivative of the comoving distance, must be inferred indirectly when starting from raw SN~Ia data.

A variety of interrelated methods have been used for this purpose. Some analyses have focused on model-independent reconstruction of an analytical $E(z)$ function or of other dynamical quantities like the deceleration parameter $q(z)$ \citep{Shafieloo:2005nd,Sahni:2006pa,Shafieloo:2007cs,Ishida:2010nk}. Such reconstructions are useful for understanding where the data are most constraining, and they can indicate whether the functional form for $H(z)$ naturally preferred by the data is consistent with that of a physical model like $\Lambda$CDM. On the other hand, it is not possible, or at least not straightforward, to subsequently incorporate the reconstructions in a likelihood function, or otherwise in a statistical analysis, in order to constrain cosmological parameters.

Other methods focus on obtaining direct measurements of $E(z)$ at several redshifts by smoothing and/or weighting the individual SNe and differentiating the distance-redshift relation \citep{Tegmark:2001zc,Daly:2003iy,Daly:2004gf}. One proposed method \citep{Wang:2005yaa}, which has been employed in some subsequent analyses \citep{Riess:2007,Mortsell:2008yu,Avgoustidis:2009ai}, seeks direct, independent estimates of $E(z)$ in redshift bins by first converting SN distance moduli into their corresponding comoving distances $r_i$, then transforming these $r_i$ into noisy, but locally unbiased, estimates of $E(z)^{-1}$ between neighboring SNe. A specific weighted average then yields a minimum-variance estimate of $E(z)^{-1}$ over a wider redshift bin. We have verified numerically that this procedure is actually \emph{equivalent} to the weighted least-squares fit of a line to the $r_i$ vs.\ $z_i$ data over the same wide redshift bin, where the slope corresponds to $E(z)^{-1}$. Both the least-squares estimator and that of \cite{Wang:2005yaa} have been shown to be unbiased and have minimum variance, assuming SN redshifts are exact and $E(z)$ is constant over the redshift bin, so it is not surprising that these estimators coincide.

While such an approach is attractive in that it directly transforms the SN distances into independent measurements of $E(z)$ at different redshifts, it has notable problems that make it unsuitable in practice. The first step requires converting SN distance moduli into comoving distances, and one must therefore assume a value for the intercept of the Hubble diagram, which is unknown \textit{a priori}. As this quantity is partially degenerate with $E(z)$, particularly the lowest-redshift measurement, fixing the intercept to some best-fit value would artificially remove a degree of freedom from the fit, resulting in underestimated uncertainties. One could instead interpret the estimates as estimates of $A E(z)^{-1}$, where $A$ is an arbitrary constant. In this case, though, properly extracting cosmological information from the $E(z)$ measurements would require fully marginalizing over $A$ in a fit to multiple measurements of $A E(z)^{-1}$.

Furthermore, an $E(z)$ estimate using this method reflects some average of $E(z)$ over the redshift bin, not necessarily the value at the bin's center. Unless $E(z)$ is constant over the redshift bin, this will lead to a bias, and since only a handful of $E(z)$ values can be constrained robustly with current data, one might expect the bias to be significant. Indeed, by simulating instances of our SN data (see Section~\ref{sec:interpfits}), we have verified that biases in such $E(z)$ estimates are typically a large fraction ($\sim\!0.5$) of their uncertainty, making the measurements unsuitable for later cosmological inference.

\subsection{Parametrized \texorpdfstring{$E(z)$}{E(z)} and Interpolation} \label{sec:interpfits}

We now describe a somewhat different approach for determining $E(z)$ from SN~Ia data. We explain how it avoids the problems discussed in Section~\ref{sec:methods} and provides more meaningful and robust $E(z)$ measurements. We will assume that the true, underlying $E(z)$ function is a continuous, smooth function of redshift, which is certainly the case for most physical and empirical models studied in the literature.

In our approach, we parametrize $E(z)$ by its value at several specific redshifts and employ a basic interpolation scheme to define the complete $E(z)$ function, which can then be numerically integrated to compute the luminosity distance and compare to the data. This allows us to constrain the $E(z)$ parameters using the full SN dataset in its raw form, as one would in a standard dark energy analysis. This way, any nuisance parameters associated with the SN data, notably the distance scale or Hubble diagram intercept, can be properly marginalized over in the fit.

While the total number of $E(z)$ values to constrain is somewhat arbitrary, there are several considerations. Choosing too many $E(z)$ parameters results in weaker constraints and posterior distributions that are less likely to be Gaussian. Choosing too few $E(z)$ values increases the chance that the estimates will be biased, as the interpolating function will deviate from the functional form of the underlying cosmology. The specific redshifts, while also somewhat arbitrary, should reflect the redshift range and distribution of the SNe. Since the $E(z)$ measurements, especially those at neighboring redshifts, will naturally be somewhat correlated, choosing too small a separation in redshift between a given pair will lead to undesirably large pairwise correlations in the estimates.

Overall, we find that employing a shape-preserving piecewise-cubic Hermite interpolating polynomial (implemented as \texttt{pchip} in MATLAB; see \citealt{NMSbook}) to interpolate (and extrapolate) the $E(z)$ function works particularly well, though other interpolation schemes (various splines, simple linear interpolation) are also generally suitable. For any specified $E(z)$ (any fiducial cosmology), it is straightforward to determine whether the $E(z)$ estimates resulting from the interpolation and fitting procedure are unbiased. To check this, we repeatedly simulate instances of our SN~Ia data; that is, we keep the same SN redshifts and covariance matrix as the real SN data, but repeatedly sample the distance moduli from a multivariate Gaussian centered on the fiducial cosmology. Of course, unbiased constraints for the fiducial cosmology do not guarantee unbiased results for other cosmologies. In principle, one could perform this check for each specific model of interest; however, there is reason to worry only when a model predicts $E(z)$ to vary rapidly or have features too narrow to be captured by the widely spaced $E(z)$ parameters. For the highest-redshift SNe~Ia, a modest amount ($\sim\!25$\%) of the integral of $E(z)^{-1}$ must be evaluated via extrapolation beyond the last redshift anchor of the $E(z)$ function. However, our simulations indicate that this does not bias this highest-redshift measurement of $E(z)$. Indeed, we have verified that all of the $E(z)$ measurements are biased by $\lesssim 10$\% of their individual statistical uncertainties.

In essence, our procedure trades the ability to make direct, independent measurements of $E(z)$ at redshifts that are somewhat uncertain (and not randomly so) for the ability to obtain precise, unbiased $E(z)$ estimates at specific redshifts. Only this latter type of estimate allows for accurate subsequent cosmological inference with the $E(z)$.

\begin{figure*}[h]
\centering
\includegraphics[width=0.8\textwidth]{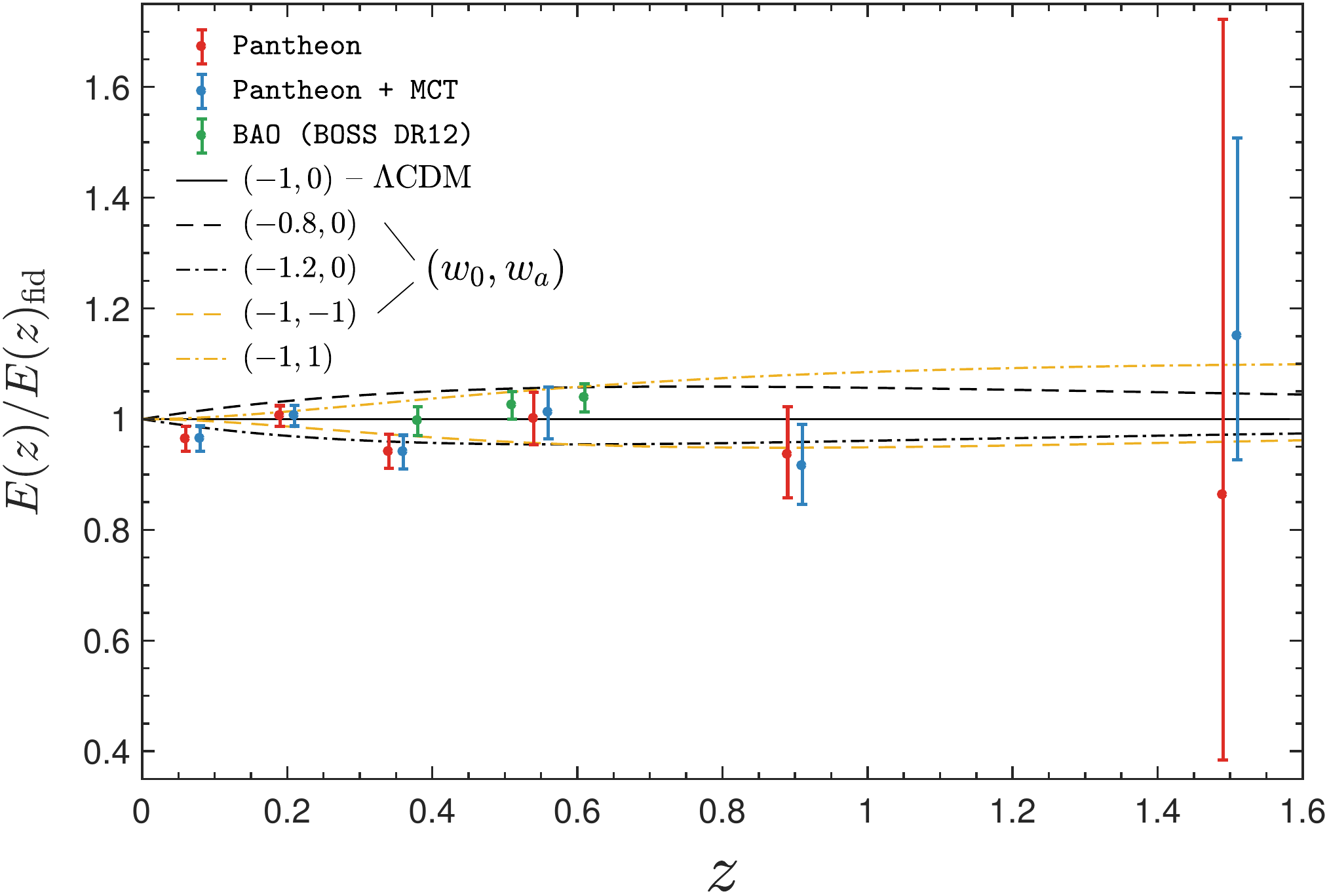}
\caption{Constraints on $E(z) \equiv H(z)/H_0$, relative to $E(z)$ for a fiducial $\Lambda$CDM model ($\Omega_m = 0.3$). We compare the constraints with (blue points) and without (red points) the high-redshift CANDELS and CLASH (MCT) SNe~Ia. Note that these $E(z)$ measurements are correlated and have non-Gaussian distributions (the error bars enclose 68.3\% of the likelihood). For comparison, we also show the three (correlated) measurements of $E(z)$ from combined BOSS DR12 BAO data \citep{Alam:2016hwk} after calibration with \textit{Planck} $\Lambda$CDM constraints on $H_0 \, r_d$ (green points).}
\label{fig:Hz}
\end{figure*}

\begin{figure*}[h]
\centering
\includegraphics[width=0.8\textwidth]{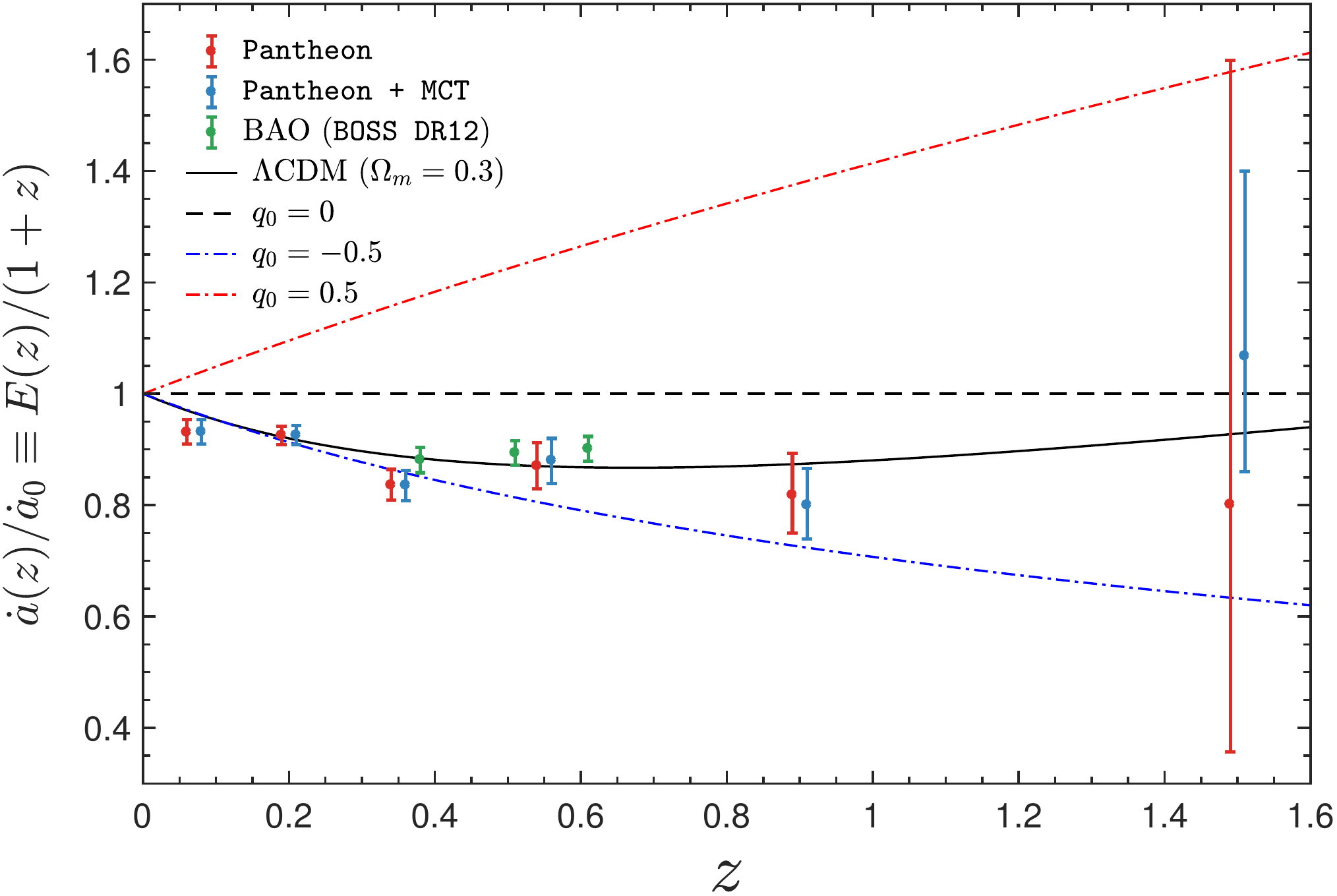}
\caption{For the same data as in Figure~\ref{fig:Hz}, we show constraints on the time derivative of the scale factor $\dot{a}(z)$ relative to its present value, obtained by scaling the $E(z)$ values by $(1 + z)^{-1}$. We compare the fiducial $\Lambda$CDM model to alternative models with a constant deceleration parameter $q_0 = 0$ (coasting cosmology), $q_0 = -0.5$ (pure acceleration), and $q_0 = 0.5$ (pure deceleration), all assuming a flat universe.}
\label{fig:adot}
\end{figure*}

\begin{figure*}[h]
\centering
\includegraphics[width=0.46\textwidth]{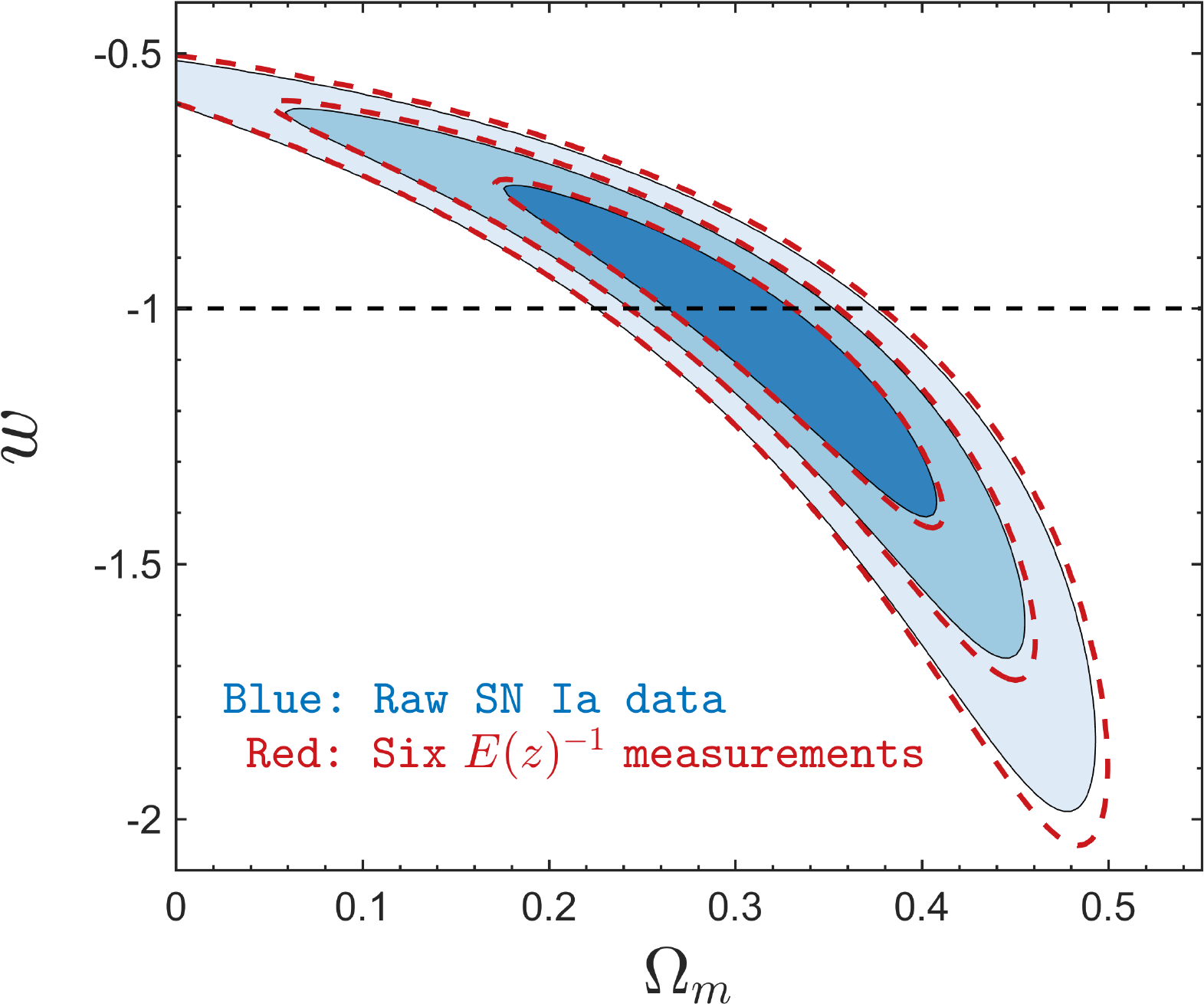}
\hspace{0.05\textwidth}
\includegraphics[width=0.45\textwidth]{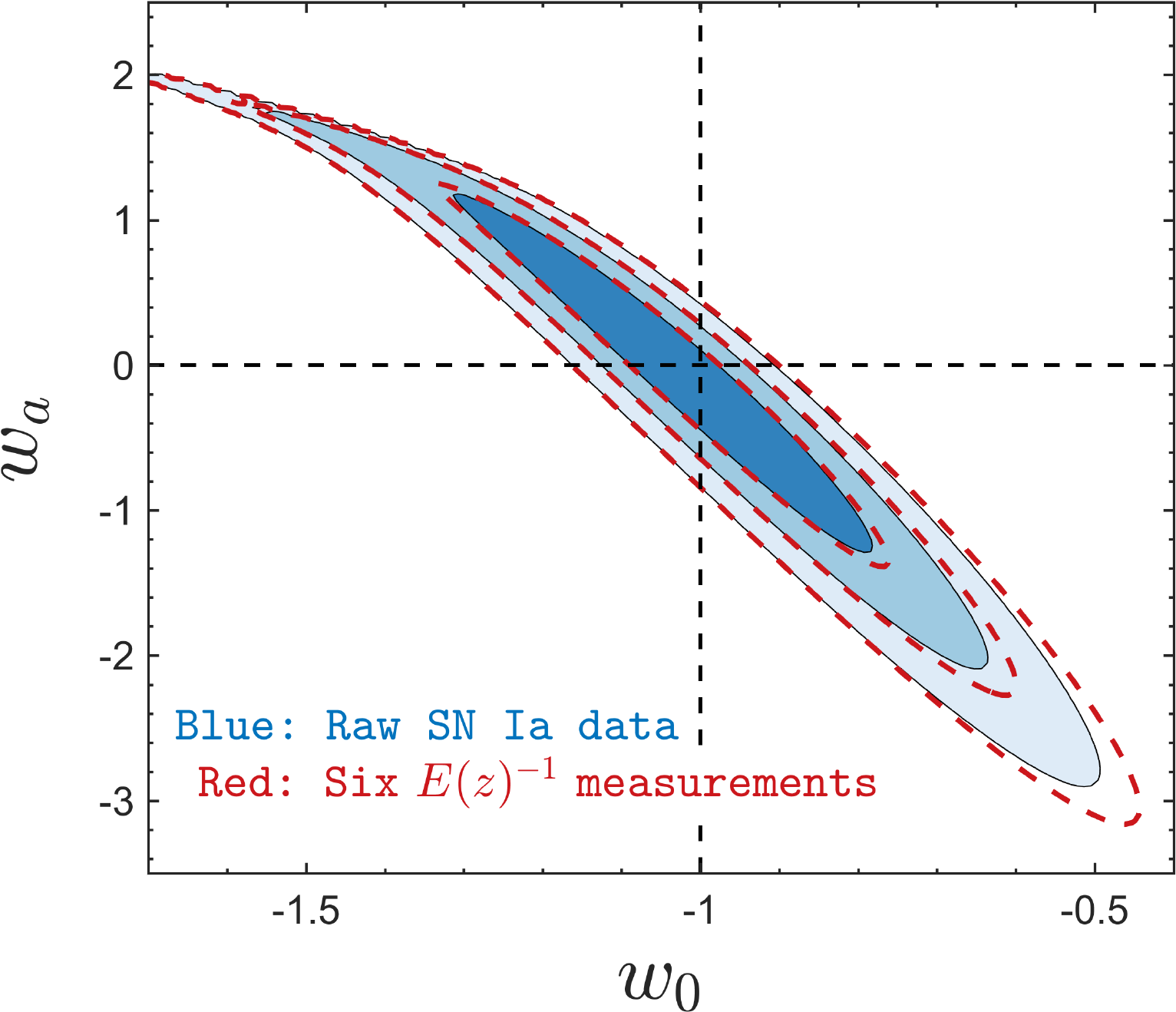}
\caption{Constraints on $\Omega_m$ and a constant equation-of-state parameter $w$ in a flat universe (left panel) and for the $w_0$--$w_a$ model \citep{Chevallier:2000qy,Linder:2002et}, marginalized over $\Omega_m$ and also assuming a flat universe (right panel). We compare the constraints when using the full SN~Ia likelihood with individual distance moduli (filled blue contours) with the constraints from the six moderately correlated $E(z)$ measurements (open red contours). Contours contain 68.3\%, 95.4\%, and 99.7\% of the likelihood, and for the $w_0$--$w_a$ constraints we have also included distance priors derived from \textit{Planck} data \citep{Ade:2015xua}.}
\label{fig:DE_compare}
\end{figure*}

\subsection{SN Ia Constraints on \texorpdfstring{$E(z)$}{E(z)}} \label{sec:results}

\begin{deluxetable*}{ccrrrrrrcc}
\tablecaption{Pantheon + MCT SN~Ia Measurements of $E(z)$ \label{tab:Ez}}
\tablehead{\colhead{$z$} & \colhead{$E(z)^{-1}$ \tablenotemark{a}} & \multicolumn{6}{c}{Correlation Matrix} & \colhead{$E(z)$} & \colhead{Distance Residual $\Delta \mu / (0.01~\text{mag}$) \tablenotemark{b}}}
\startdata
0.07 & 1.003 $\pm$ 0.023 &    1.00 &         &         &         &         &      & 0.997 $\pm$ 0.023      & $-$0.13 $\pm$ 0.99 \\
 0.2 & 0.901 $\pm$ 0.017 &    0.39 &    1.00 &         &         &         &      & 1.111 $\pm$ 0.020      & $-$0.23 $\pm$ 1.26 \\
0.35 & 0.887 $\pm$ 0.029 &    0.53 & $-$0.14 &    1.00 &         &         &      & 1.128 $\pm$ 0.037      & $+$0.23 $\pm$ 1.32 \\
0.55 & 0.732 $\pm$ 0.033 &    0.37 &    0.37 & $-$0.16 &    1.00 &         &      & 1.364 $\pm$ 0.063      & $+$0.11 $\pm$ 1.97 \\
 0.9 & 0.656 $\pm$ 0.052 &    0.01 & $-$0.08 &    0.17 & $-$0.39 &    1.00 &      & 1.52  $\pm$ 0.12       & $+$1.15 $\pm$ 2.85 \\
 1.5 & 0.342 $\pm$ 0.079 & $-$0.02 & $-$0.08 & $-$0.07 &    0.15 & $-$0.19 & 1.00 & $2.67^{+0.83}_{-0.52}$ & $-$3.42 $\pm$ 6.78 \\
\enddata
\tablenotetext{a}{Mean and standard deviation of the marginalized likelihood, approximately Gaussian in all cases.}
\tablenotetext{b}{Effective distance moduli relative to those of a fiducial $\Lambda$CDM cosmology ($\Omega_m = 0.3$), as determined by an interpolated fit to the residuals using the same redshift control points as the $E(z)$ analysis.}
\end{deluxetable*}

We now constrain $E(z)$ for the Pantheon compilation of 1039 SNe~Ia, which we will supplement with the high-redshift CANDELS and CLASH SNe. The Pantheon compilation \citep[][in prep.]{Scolnic:2017} includes data from multiple surveys (CfA(1--4), CSP, SDSS, SNLS, Pan-STARRS1, \textit{HST}) calibrated for a joint cosmological analysis. Below we summarize the key aspects of the Pantheon analysis, and we refer the reader to \cite{Scolnic:2017} for additional details and a complete discussion.

The Pantheon analysis presents the full set of spectroscopically confirmed SNe~Ia from the Pan-STARRS1 (PS1) Medium Deep Survey, building on the earlier analysis of the first 1.5~yr of PS1 \citep{Rest:2014,Scolnic:2013efb}. It relies on the Supercal cross-calibration procedure presented by \cite{Scolnic:2015eyc}, which uses the relative consistency of the Pan-STARRS1 photometry over 3$\pi$ steradians of the sky to tie together the photometric systems of the individual surveys. The Pantheon analysis also incorporates the BBC methodology of \cite{Kessler:2016uwi} \citep[see also][]{Scolnic:2016ukm}, which corrects for distance biases dependent on the light-curve properties of the SNe and the surveys from which they are selected.

The Pantheon analysis employs the SALT2 light-curve fitter \citep{Guy:2007dv,Betoule:2014}, which determines an overall normalization of the log-flux ($m_B$), a shape parameter ($x_1$), and a color ($c$) for each SN light curve, along with associated uncertainties. We standardize the SNe by modeling an individual SN~Ia distance modulus as
\begin{equation}
\mu = m_B - M + \alpha \, x_1 - \beta \, c + \Delta_M + \Delta_B \, .
\end{equation}
The $\Delta_M$ term is an additional correction for the empirical host-mass step, where SNe in high-stellar-mass host galaxies ($\log_{10}(M_\ast/{\rm M}_\odot) \gtrsim 10$) are $\sim\!0.05$~mag brighter on average, after light-curve standardization. The $\Delta_B$ term represents the distance bias correction. Note that $M$, $\alpha$, $\beta$, and the amplitude of the mass step (included in the $\Delta_M$ term) are all nuisance parameters that must be determined by a fit to the data. In our analysis, only $M$ (effectively, the Hubble diagram offset) is fit along with the cosmological parameters $E(z)$. The other parameters are well determined independently of cosmology in the Pantheon analysis. The inferred values are $\alpha \approx 0.15$--0.16 and $\beta \approx 3.0$--3.7, where the results vary depending on the intrinsic scatter model\footnote{In the Pantheon analysis, two alternative models for the intrinsic scatter are separately used to derive distance bias corrections, which are then averaged, with half of the difference included in the systematic uncertainty budget.}. Finally, note that the distance modulus as predicted by the cosmological model is given by
\begin{equation}
\mu(z) = 5 \log_{10} \left[\frac{d_L(z, \mathbf{p})}{1~\text{Mpc}} \right] + 25 \, ,
\end{equation}
where $d_L$ is the luminosity distance, which is a function of redshift and also depends on the set of cosmological parameters $\mathbf{p}$.

The statistical uncertainties of SN distance moduli are modeled, in the standard way, as a combination of observational measurement uncertainty, intrinsic scatter, and additional scatter due to gravitational lensing, peculiar velocities, and redshift measurement uncertainty\footnote{Separate from standard propagation of redshift uncertainty, the derived distance moduli themselves depend on the observed redshift. We have verified that, for SN GND12Col, which has a large redshift uncertainty with asymmetric errors, repeating the analysis with both its redshift and distance shifted by 1$\sigma$ does not significantly affect the results.}. The inferred value for the intrinsic scatter is $\sigma_\text{int} \approx 0.1$, although, like $\alpha$ and $\beta$, the value depends on the intrinsic scatter model. After bulk-flow corrections are applied to the low-redshift SNe, we add a peculiar-velocity scatter of $\sigma_v = 250$~km~s$^{-1}$. We assume a value $\sigma_\text{lens} = 0.055z$ for the lensing scatter \citep{Jonsson:2010b}. Note that the distribution of the shift in observed magnitude due to lensing is non-Gaussian \citep[e.g.,][]{Jonsson:2006}, with a tail of strongly magnified SNe; however, by examination of foreground structures we have verified that none of our CANDELS or CLASH SNe are likely to fall in this tail, making the lensing scatter contribution to the distance uncertainty a good approximation. Note that there is also statistical uncertainty in the host-mass correction and the distance bias correction.

The Pantheon analysis also includes a rigorous analysis of systematic errors, adding terms to the covariance matrix of SN distances to account for uncertainties in photometric calibration (including terms for individual survey calibration, the Supercal cross-calibration procedure, and the SALT2 model itself), the intrinsic scatter model, survey selection functions, Milky Way dust extinction, $\beta$ evolution, the host mass step and its evolution, and peculiar velocity coherent flow corrections.

Standard data-quality cuts were applied to remove SNe that are not expected to follow the empirical standardization relations. Specifically, we keep only SNe with $|x_1| < 3$, $\sigma_{x_1} < 1$, $|c| < 0.3$, a light-curve fit with $\chi^2/N_\text{dof} < 3$, and an uncertainty in the time of peak brightness of less than 2 days. Similar cuts have been used in most recent SN~Ia cosmological analyses \citep[e.g.,][]{Betoule:2014,Rest:2014,Riess:2016jrr}. These cuts eliminate 3 of the silver and gold MCT SNe (CLH11Tra, GND13Gar, GND13Jay; see Table~\ref{tab:distances}). Finally, a $4\sigma$ outlier rejection from the best-fit Hubble diagram is applied and removes GND13Cam, leaving 9 \textit{HST} MCT SNe~Ia in the joint analysis\footnote{In the Pantheon analysis, additional cuts were applied to remove SNe without an observation at least 5~days after peak brightness and with light-curve parameters that do not fall in the simulated distribution from the BBC method \citep[see][]{Scolnic:2017}. These cuts do not remove any of the remaining MCT SNe.}. Note that here we do include EGS13Rut, which is on the edge of the $\sigma_{x_1}$ cut but has typical light-curve fit parameters. Although the final MCT addition of 9 SNe represents $<\!1$\% of the combined sample, the unusually-high redshifts (7 with $z > 1.5$) provide unique leverage on $E(z)$ at $z = 1.5$.

Following the methodology and discussion in Section~\ref{sec:interpfits}, we parametrize $E(z)^{-1}$ by its value at six redshifts (chosen to best summarize the sample) and therefore have six free parameters to constrain. It is important to remember that the Hubble diagram offset is a free parameter as well, though we analytically marginalize over this offset, with a flat prior, in the likelihood. We assume a flat universe ($\Omega_k = 0$) throughout, so the $E(z)$ measurements are cosmological-model-dependent in this sense. To obtain the constraints, we sample the likelihood using a custom Markov chain Monte Carlo (MCMC) code employing the basic Metropolis-Hastings algorithm. We impose flat, hard-bound priors on the $E(z)^{-1}$ parameters wide enough that extending the bounds does not affect the resulting constraints. The final MCMC chains were inspected to verify convergence.

The resulting marginalized posterior likelihoods for $E(z)^{-1}$ are Gaussian to a good approximation, and the constraints are given in Table~\ref{tab:Ez}. In Figure~\ref{fig:Hz}, we convert the measurements of $E(z)^{-1}$ into $E(z)$ measurements by reprocessing the MCMC chains and then compare the results with and without the MCT SNe. It is not surprising that the MCT SNe subsantially improve the measurement of $E(z)$ at $z = 1.5$. They permit a $\sim\!20$\% measurement of $E(z = 1.5)$, roughly a factor of three improvement over the result without the MCT SNe. While the CANDELS and CLASH SNe mostly affect the measurement at $z = 1.5$, they also improve and shift some lower-redshift measurements, which are somewhat correlated ($\approx\!8$\% and 4\% improvements at $z = 0.9$ and 0.55, respectively). By eye, the set of $E(z)$ measurements may appear somewhat discrepant with the fiducial $\Lambda$CDM model, but the overall $\chi^2$, which includes the moderate correlations, is 5.6 for the 6 degrees of freedom.

In Figure~\ref{fig:adot}, we scale $E(z)$ by $(1 + z)^{-1}$ to illustrate the constraints on the time derivative of the scale factor $\dot{a}(z)$, relative to its present value, for the same data shown in Figure~\ref{fig:Hz}. In this space, it is clear that the low-redshift and high-redshift $E(z)$ measurements together provide evidence for both recent acceleration and earlier deceleration epochs, as predicted by standard cosmological models. In addition to the fiducial $\Lambda$CDM model, we show dynamical models with fixed deceleration parameter $q_0$. The $\dot{a}(z)$ values track the $q_0 = -0.5$ model at $z \lesssim 0.5$ (where the low-$z$ behavior matches that of a $\Lambda$CDM model with $\Omega_m \approx 0.3$) but show deceleration with respect to that curve at higher redshifts. The coasting cosmology ($q_0 = 0$), pure acceleration cosmology ($q_0 = -0.5$), and pure deceleration cosmology ($q_0 = 0.5$, equivalent to a flat CDM model with $\Omega_m = 1$) are strongly disfavored with $\Delta \chi^2 = 79.8$, $\Delta \chi^2 = 36.5$, and $\Delta \chi^2 = 360.2$, respectively, for 6 degrees of freedom. The measurement at $z = 1.5$ alone, while consistent with the other models, disfavors $q_0 = -0.5$ with $\Delta \chi^2 = 13.5$.

As an illustration of the power of the $E(z)$ measurements in constraining (spatially flat) cosmologies, we compare constraints on common dark energy parameterizations in Figure~\ref{fig:DE_compare}. Remarkably, the constraints are nearly identical whether the parameters are constrained with the SN~Ia data directly or with the $E(z)$ measurements in Table~\ref{tab:Ez}. It may not be too surprising that $E(z)$ captures the constraining power of the SNe for simple one-or-two-parameter models. One would not expect the same for fits with many degrees of freedom (e.g., more complicated dark energy models); in practice, however, current and near-future SN~Ia data can only meaningfully constrain 2--3 expansion parameters anyway. For models that assume a flat universe and predict fairly smooth, featureless $H(z)$, the $E(z)$ constraints will be an efficient summary of the present SN~Ia data.

\subsection{High-Redshift SNe Ia and Evolution} \label{sec:SN_evol}

The use of SNe~Ia as standardizable candles across redshift relies on the understanding that their uncommonly homogeneous luminosities and colors follow from their nature as carbon-oxygen white dwarfs close to the Chandrasekhar mass. While uncertainty persists regarding \emph{how} these degenerate stars approach that mass limit, either by accretion from a nondegenerate companion or through the tidal disruption followed by accretion of a degenerate companion, there has long been agreement about this model based on the well-understood physics of degenerate stars \citep{Hoyle:1960zz,Arnett1969,ColgateMcKee1969}. The thermonuclear detonation of a Chandrasekhar-mass carbon-oxygen white dwarf yields a mass of radioactive nickel whose energy output matches that of a SN~Ia \citep{ArnettBranchWheeler1985} and whose modeled nucleosynthesis matches its spectral elements \citep{Nomoto:1984sm}. More recently, prediscovery observations of SN~2011fe, a prototypical SN~Ia in M101, demonstrated that the progenitor did not exceed a radius of 2\% solar, fully consistent with the expected white dwarf \citep{Nugent:2011cz,Li:2011nv,Bloom:2011au}. Yet the difficulty and low likelihood of ever directly observing a white dwarf system before it becomes a SN~Ia leaves enough uncertainty and model freedom to support the consideration of redshift evolution of the standardized SN~Ia luminosity.

From SN~Ia observations spanning a wide range of redshifts and sampling the epochs when cosmic expansion accelerates and decelerates, it is possible to distinguish such evolution from the uncertain properties of dark energy \citep{Riess:2006yk}. As an illustration of the power of SNe~Ia at $z > 1$ to separate evolution from cosmology, we briefly reconsider the analysis of \cite{Tutusaus:2017ibk}, which shows that power-law cosmology, where the scale factor evolves as $a(t) \propto t^n$ for some exponent $n$, is an equally good fit to SN~Ia data (primarily at $z < 1$) as the $\Lambda$CDM model (with $\Omega_m$ free) when the standardized luminosity is also allowed to vary with redshift according to some simplistic, empirical models of SN~Ia evolution. Although such models are not astrophysically motivated, they may be useful for exploring the separation of other SN distance-dependent effects (e.g., grey extinction) from cosmological parameters.

\begin{figure*}[h]
\centering
\includegraphics[width=0.75\textwidth]{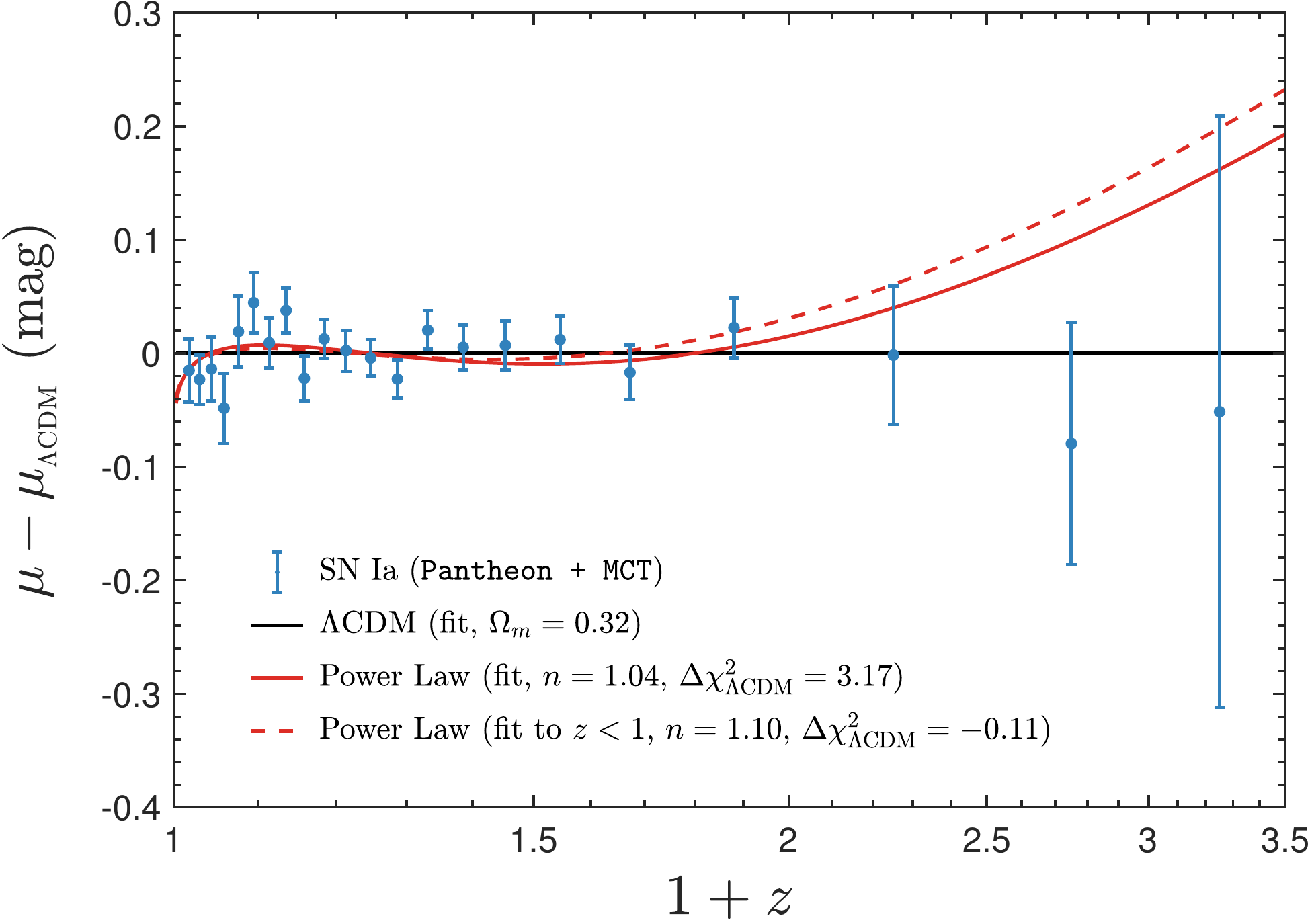}
\caption{Comparison of $\Lambda$CDM and power-law cosmology ($a(t) \propto t^n$) fits to our SN~Ia data, where in each case we allow the intrinsic luminosity to evolve as $\Delta M(z) = \epsilon z^\delta$, corresponding to Model B from \cite{Tutusaus:2017ibk}, where we fix $\delta = 0.3$. The SN data are binned for clarity, and $\Delta \chi^2_{\Lambda \text{CDM}} \equiv \chi^2 - \chi^2_{\Lambda \text{CDM}}$.}
\label{fig:SN_evol}
\end{figure*}

Here, as an illustration, we consider Model B ($\Delta M(z) = \epsilon z^\delta$) from \cite{Tutusaus:2017ibk} with fixed $\delta = 0.3$. We separately fit both $\Lambda$CDM and power-law cosmology to our combined (Pantheon + MCT) data; in each case, we fit for the Hubble diagram intercept, a cosmological parameter ($\Omega_m$ or $n$), and the amplitude $\epsilon$ of the assumed intrinsic luminosity evolution. We compare these fits in Figure~\ref{fig:SN_evol}. Fitting only the SNe at $z < 1$, a power law with $n = 1.1$ is a slightly \emph{better} fit to the SN~Ia data than $\Lambda$CDM. Indeed, when analyzing the JLA compilation, which features only $\sim\!5$ SNe at $z > 1$, \cite{Tutusaus:2017ibk} claims a mild preference for the power law (note that their analysis also included BAO and $H(z)$ information).

In contrast, when we include our 24 SNe at $z > 1$, a nearly coasting (marginally accelerating) power-law cosmology (best fit $n = 1.04$) together with simplistic SN~Ia evolution is no longer as good a fit as the $\Lambda$CDM model, with a relative probability of $\exp(-\Delta\chi^2/2) \approx 20\%$. Without invoking evolution (that is, fixing $\epsilon \equiv 0$), the $\Lambda$CDM model is a much better fit than the power-law model, with the latter strongly disfavored with $\Delta \chi^2_{\Lambda \text{CDM}} = 8.3$, a relative probability of 1.6\%, when including the new SNe at $z > 1.5$. Meanwhile, assuming $\Lambda$CDM and fitting for the evolution amplitude $\epsilon$ yields a value consistent with zero, $\epsilon = 0.08 \pm 0.15$, so there is no motivation for including it based on astrophysical \emph{or} empirical considerations. A more comprehensive investigation of SN~Ia evolution and cosmology is underway \citep[][in prep.]{Shafer:2017}.

We note the addition of the MCT SNe to the Pantheon compilation also further reduces the already-low likelihood of the ``empty universe'' solution where $\Omega_m \approx 0$ and $\Omega_\Lambda \approx 0$ in an $O\Lambda$CDM universe, a location \cite{Nielsen:2015pga} claimed to be marginally consistent ($\sim\!3\sigma$) with SN data alone using unconventional priors on SN distributions, to the boundary of the 6$\sigma$ contour.

\subsection{\texorpdfstring{$E(z)$}{E(z)} with \textit{WFIRST}} \label{sec:wfirst}

The \textit{Wide-Field Infrared Survey Telescope} (\textit{WFIRST}) was the top space-based recommendation of the 2010 U.S. astronomy and astrophysics decadal survey. The mission is still in formulation, but current plans specify a 2.4~m primary mirror and include a wide-field instrument for cosmology. The cosmology science objectives, as detailed in the most recent report from the Science Definition Team \citep[SDT;][]{Spergel:2015sza}, will be accomplished through a combination of SN~Ia, galaxy, and weak-lensing surveys.

The \textit{WFIRST} SN survey is anticipated to yield a large sample of thousands of SNe, many at $z > 1$ with precise distances. These SNe will vastly improve upon the high-redshift $E(z)$ measurements available today, allowing nontrivial and precision tests of the $\Lambda$CDM model independent of the BAO and weak-lensing constraints in a redshift range that is currently not well constrained.

Here we wish to forecast realistic constraints on $E(z)$ from \textit{WFIRST}. Typical forecasts (e.g., for dark energy figures of merit) rely on Fisher matrix formalism, which is exact only for Gaussian posterior distributions and otherwise underestimates parameter uncertainties. For SN Ia forecasts, one typically assumes idealized, or roughly estimated, redshift distributions and makes simple assumptions about the measurement error. Here instead we employ a detailed simulation of one potential observing strategy for the \textit{WFIRST} SN survey \citep{Hounsell:2017ejq}. We then constrain the $E(z)$ parameters using the methodology of Section~\ref{sec:interpfits} that was employed in Section~\ref{sec:results} for our current Pantheon + MCT data.

For our illustration, we consider the \texttt{Imaging All-z} strategy described by \cite{Hounsell:2017ejq}. This particular strategy relies on multi-band imaging for classification and assumes follow-up spectroscopy will provide host-galaxy redshifts. \cite{Hounsell:2017ejq} also assumes a large external sample of 800 SNe at $z < 0.1$. As the size of future systematic uncertainties is hard to predict, \cite{Hounsell:2017ejq} simulates a range of scenarios, and here we opt for all-around optimistic assumptions about future systematic errors \citep[for what this entails, see][]{Hounsell:2017ejq}. In this scenario, the contribution of systematic errors is not negligible but is subdominant in the error budget.

\begin{figure*}[h]
\centering
\includegraphics[width=0.8\textwidth]{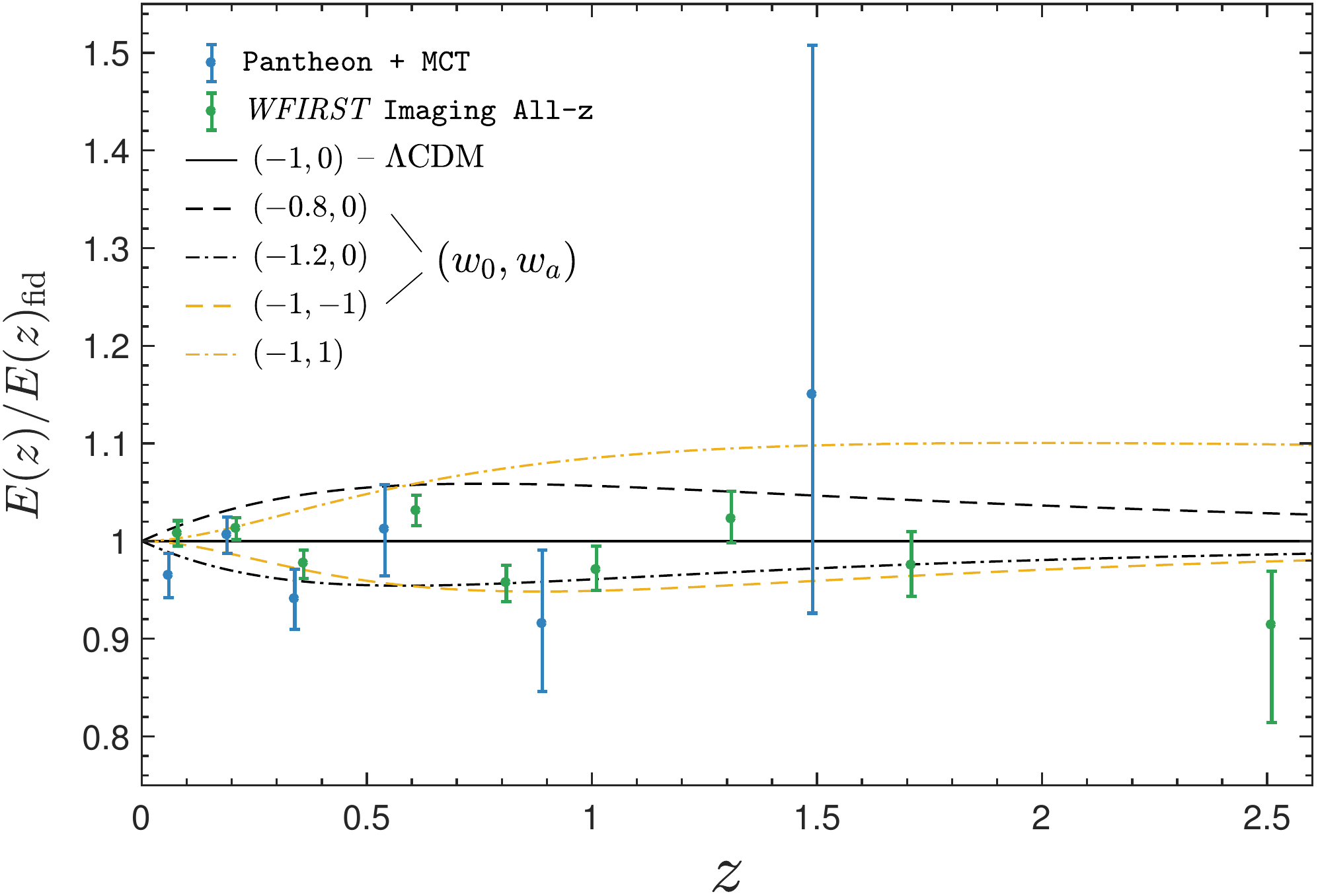}
\caption{Simulated \textit{WFIRST} constraints on $E(z) \equiv H(z)/H_0$, relative to $E(z)$ for a fiducial $\Lambda$CDM model ($\Omega_m = 0.3$). We compare the constraints from current data (blue points) with simulated constraints from the \textit{WFIRST} \texttt{Imaging All-z} observing strategy (green points). We overlay the same dark energy models as in Figure~\ref{fig:Hz}.}
\label{fig:Hz_WFIRST}
\end{figure*}

In Figure~\ref{fig:Hz_WFIRST}, we compare our current Pantheon + MCT constraints on $E(z)$ with simulated constraints from the \textit{WFIRST} \texttt{Imaging All-z} strategy. We find that we are able to constrain $E(z)$ robustly, albeit with moderate pairwise correlations, at 9 redshifts in the range $0.07 < z < 2.5$. In Table~\ref{tab:Ez_WFIRST}, we list the percent errors for $E(z)$ corresponding to our forecast. Note that these results are negligibly changed whether we quote percent errors on $E(z)$ or its inverse. We find that \textit{WFIRST} allows 8 measurements of $E(z)$ at the 1--3\% level, along with a robust but less precise measurement at $z \approx 2.5$. Notably, this is a constraint on the expansion rate at a redshift higher than any SN~Ia has even been observed to date.

\begin{deluxetable}{cc} 
\tablecaption{Simulated SN~Ia Measurements of $E(z)$ from \textit{WFIRST} \label{tab:Ez_WFIRST}}
\tablehead{\colhead{$z$} & \colhead{$E(z)$ Percent Error \tablenotemark{a}}}
\startdata
0.07 & 1.3 \\
0.2  & 1.1 \\
0.35 & 1.5 \\
0.6  & 1.5 \\
0.8  & 2.0 \\
1.0  & 2.3 \\
1.3  & 2.6 \\
1.7  & 3.4 \\
2.5  & 8.9 \\
\enddata
\tablenotetext{a}{Note that these measurements are not fully independent; there are moderate pairwise correlations among some in the set. They assume a flat universe.}
\end{deluxetable}

\section{Summary and Conclusions} \label{sec:conclude}

In this study, we analyzed the set of 15 high-redshift SNe~Ia from the CANDELS and CLASH \textit{HST} MCT programs, 9 of which ultimately pass classification confidence and quality cuts and 7 of which are at $z > 1.5$ where the relative expansion rate is poorly constrained. These are the first distance estimates for these SNe that are suitable for a joint cosmological analysis with a large compilation of lower-redshift SNe (the Pantheon compilation). We have introduced and employed a procedure to obtain unbiased constraints on the scale-free Hubble parameter $E(z) \equiv H(z)/H_0$ using only this extended Pantheon + MCT sample of SNe~Ia (Table~\ref{tab:Ez}, Figures~\ref{fig:Hz}--\ref{fig:adot}). The CANDELS and CLASH SNe at $z \gtrsim 1.5$ extend the Hubble diagram and allow us to achieve a robust measurement of the expansion rate at $z = 1.5$ that efficiently summarizes the cosmological leverage of these new SNe. Our measurement of $E(z = 1.5)^{-1} = 0.342 \pm 0.079$ (equivalently, $E(z = 1.5) = 2.67^{+0.83}_{-0.52}$) assumes a flat universe and smooth expansion history but is otherwise model-independent.

We also have demonstrated that the set of $E(z)$ measurements can serve as a form of SN~Ia data compression, allowing us to summarize SN~Ia constraints on spatially flat cosmological models that feature a smooth expansion history, which comprise the majority of the commonly studied dark energy models. The $E(z)$ are very economical, accurately reproducing parameter posteriors (even when non-Gaussian) using just 6 measured quantities in place of $>\!1000$ (Figure~\ref{fig:DE_compare}). The computation time for this $E(z)$ likelihood, relative to that for the full SN Ia likelihood, is negligible.

Future large, high-quality samples of high-redshift SNe~Ia, notably from \textit{WFIRST}, will allow precision constraints on the dark energy equation-of-state parameter $w$, especially for dynamical dark energy featuring a time-varying value of $w$. Still, there are uses for such high-redshift SNe beyond direct dark energy constraints, inspiring us to perform two additional investigations.

First, using our combined Pantheon + MCT set of SNe~Ia, we have briefly illustrated how the added leverage of our larger sample of SNe at $z > 1$, including 7 at $z > 1.5$, can help distinguish empirical SN~Ia evolution and nonstandard cosmological models from the $\Lambda$CDM model (Figure~\ref{fig:SN_evol}). We have shown that, while a nearly coasting power-law model ($a(t) \propto t^n$ with $n \approx 1$) is as good a fit to the $z < 1$ data as $\Lambda$CDM (at least when certain forms of SN evolution are allowed), adding the $z > 1$ SNe disfavors the power law, indicating a relative probability of $\sim\!20$\%, even when permitting the same SN evolution.

Second, we have used our $E(z)$ procedure in conjunction with a realistic simulation of a potential \textit{WFIRST} SN~Ia observing strategy to forecast optimistic \textit{WFIRST} constraints on $E(z)$. We find that \textit{WFIRST} will permit 8 measurements of $E(z)$ at the 1--3\% level across a wide range of redshifts, along with a robust measurement at $z \approx 2.5$ (Figure~\ref{fig:Hz_WFIRST}, Table~\ref{tab:Ez_WFIRST}). Such measurements will constitute precise tests of our expectations from the $\Lambda$CDM model separately from BAO and other high-redshift distance probes.

\bigskip

\textbf{Acknowledgments:}

We thank Rebekah Hounsell for providing the \textit{WFIRST} SN Ia simulations. It is our pleasure to thank program coordinators Patricia Royle and Beth Perriello, as well as the entire Space Telescope Science Institute (STScI) scheduling team, for their tireless efforts that made the CANDELS survey and the SN follow-up program possible.

This work was principally based on observations made with the NASA/ESA \textit{Hubble Space Telescope}, which is operated by the Association of Universities for Research in Astronomy (AURA), Inc., under NASA contract NAS5-26555. These observations are associated with program IDs 12060, 12061, 12062, 12442, 12443, 12444, 12445, 12099, 12461, and 13063. The analysis presented here made extensive use of the Mikulski Archive for Space Telescopes (MAST). STScI is operated by AURA, Inc., under NASA contract NAS5-26555. Support for MAST for non-\textit{HST} data is provided by the NASA Office of Space Science via grant NNX13AC07G and by other grants and contracts. Some of the data presented herein were obtained at the W.M. Keck Observatory, which is operated as a scientific partnership among the California Institute of Technology, the University of California, and NASA; the Observatory was made possible by the generous financial support of the W.M. Keck Foundation.

Financial support was broadly provided by NASA through grants HST-GO-12060 and HST-GO-12099 from STScI, and to S.A.R. through grant HST-HF-51312. A.V.F. is also grateful for generous financial assistance from the Christopher R. Redlich Fund, the TABASGO Foundation, and the Miller Institute for Basic Research in Science (U.C. Berkeley). A.M. acknowledges the financial support of the Brazilian funding agency FAPESP (Postdoc fellowship, process number 2014/11806-9). O.G. is supported by an NSF Astronomy and Astrophysics Postdoctoral Fellowship under award AST-1602595. J.H. was supported by a VILLUM FONDEN Investigator grant (project number 16599). S.J. was supported by JPL RSAs 143563, 1448524, 1460278, and 1473597.

\textit{Facilities:} \facility{HST (WFC3)}

\bigskip

\bibliographystyle{apj}
\bibliography{ccc}

\end{document}